\documentclass[aps,prl,twocolumn,secnumarabic,balancelastpage,amsmath,amssymb,floatfix,superscriptaddress]{revtex4-1}
\usepackage{blindtext}
\usepackage{lipsum}
\usepackage{graphics}
\usepackage{amsmath}
\usepackage{amssymb}

\usepackage{graphicx}
\usepackage{graphics}

\usepackage{verbatim}
\usepackage{physics}
\usepackage{float}
\usepackage{hyperref}
\usepackage[dvipsnames]{xcolor}
\usepackage{amsfonts}
\usepackage{braket}
\usepackage{epstopdf}
\usepackage{natbib}
\usepackage{epsfig}
\usepackage[toc,page,title,titletoc,header]{appendix}

\usepackage{dsfont,amsthm,amsbsy}
\usepackage{todonotes}
\usepackage{hyperref}
\usepackage{cancel}

\usepackage[normalem]{ulem}

\makeatletter

\let\llangle\@undefined
\let\rrangle\@undefined

\makeatother

\newcommand{\figOne}{
  \begin{figure}[t]
    \includegraphics[width=3.4in]{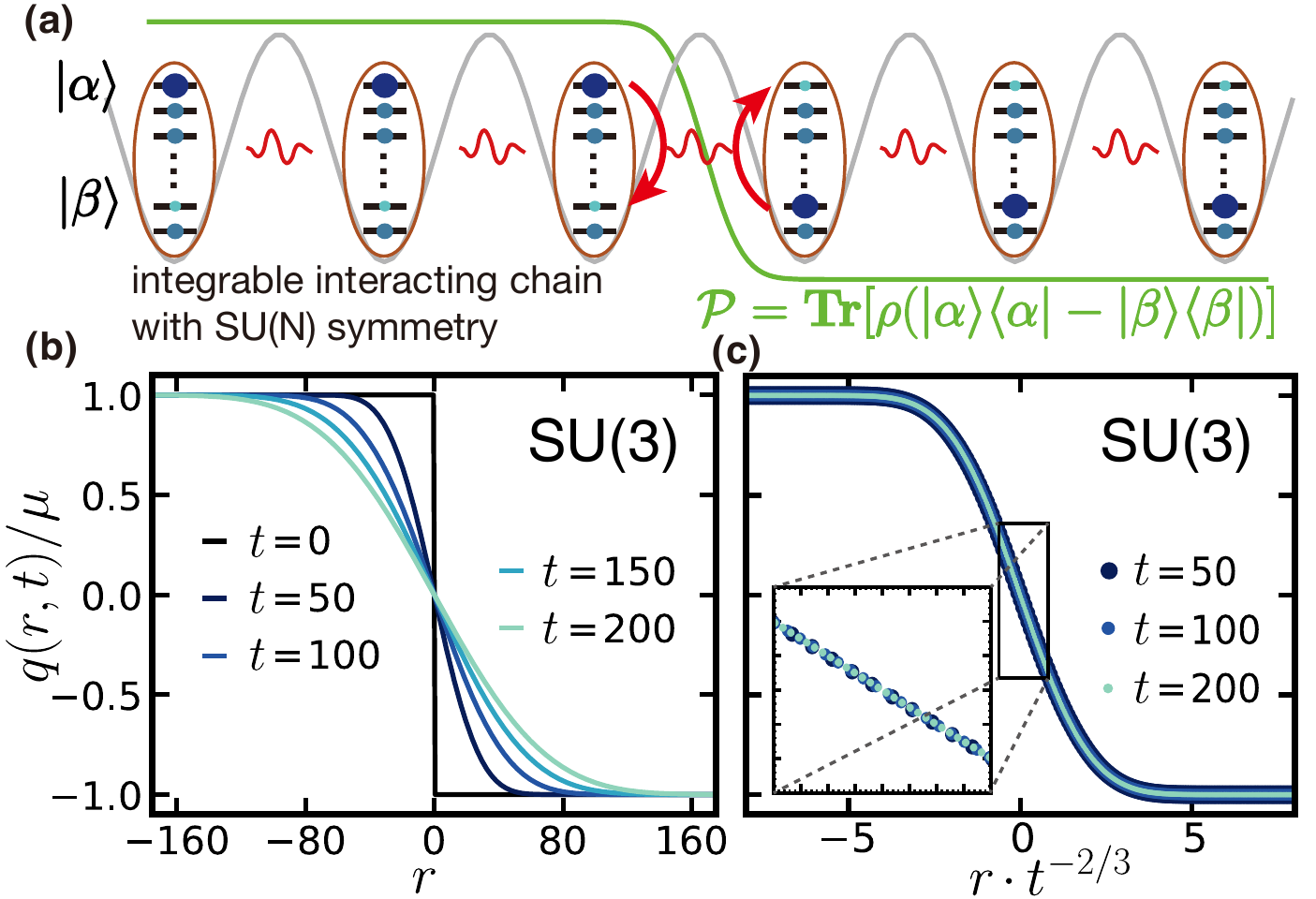}
    \centering
    \caption{
    (a) Schematic depicting a one dimensional chain of alkaline-earth atoms (each with $N$-levels) trapped in an optical lattice and interacting via nearest-neighbor super-exchange.
    The equilibration of an initial domain-wall-like imbalance encodes the underlying KPZ dynamics. 
    (b) Domain-wall  dynamics as a function of time for an SU(3)-symmetric, integrable spin chain. 
    (c) The polarization profiles at different times collapse upon rescaling with $t^{-1/z}$. The  dynamical exponent, $z=3/2$, indicates superdiffusion and is consistent with KPZ transport. 
    }
    \label{fig1}
  \end{figure}
}

\newcommand{\figTwo}{
  \begin{figure}[t]
    \includegraphics[width=3.4in]{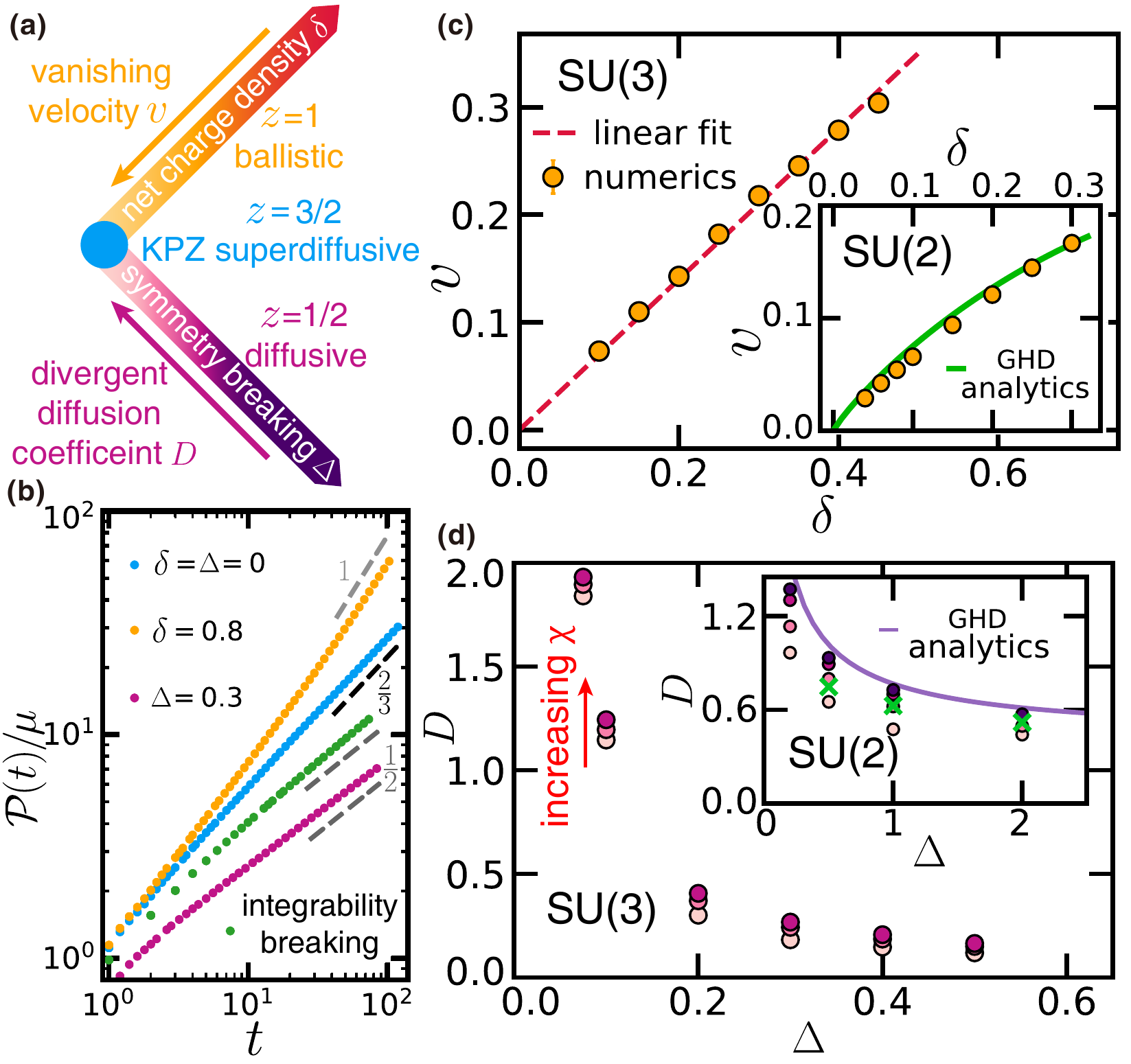}
    \centering
    \caption{
    (a) Conjectured landscape of KPZ transport in integrable, non-Abelian-symmetric models (blue dot). %
    The non-Abelian symmetry can be broken in two distinct ways, either by adding a finite charge density to the initial state (orange line) or by perturbing the underlying Hamiltonian (purple line). 
    (b) The total polarization  transferred across the domain wall, $\mathcal{P}(t)$,  directly determines the dynamical exponent. 
    For the integrable SU(3) model, $z=3/2$; when either the integrability or the symmetry is broken in the Hamiltonian, $z=2$; when the initial state has non-zero charge density,  $z=1$. 
    Note that the curve for the integrability breaking case (green) is shifted down for clarity. 
    (c) Depicts the charge transport velocity $v$ as a function of charge density $\delta$ for both the SU(3) model and the SU(2) model (inset). 
    (d) The diffusion coefficient, $D$, diverges as the integrable model approaches the SU(3)  and  SU(2) (inset) symmetric points. The DMT bond dimension, $\chi$, is chosen to be $\{64,128,256\}$ and $\{64,128,256,512\}$ for the SU(3) and  SU(2) cases, respectively. 
    Green crosses in the inset mark  previous numerical results obtained from tDMRG simulations with bond dimension $\chi\sim 2000$~\cite{karrasch2014real}.
    }
    \label{fig2}
  \end{figure}
}

\newcommand{\figThree}{
 \begin{figure*}[t]
    \centering
    \includegraphics[width=7.2in]{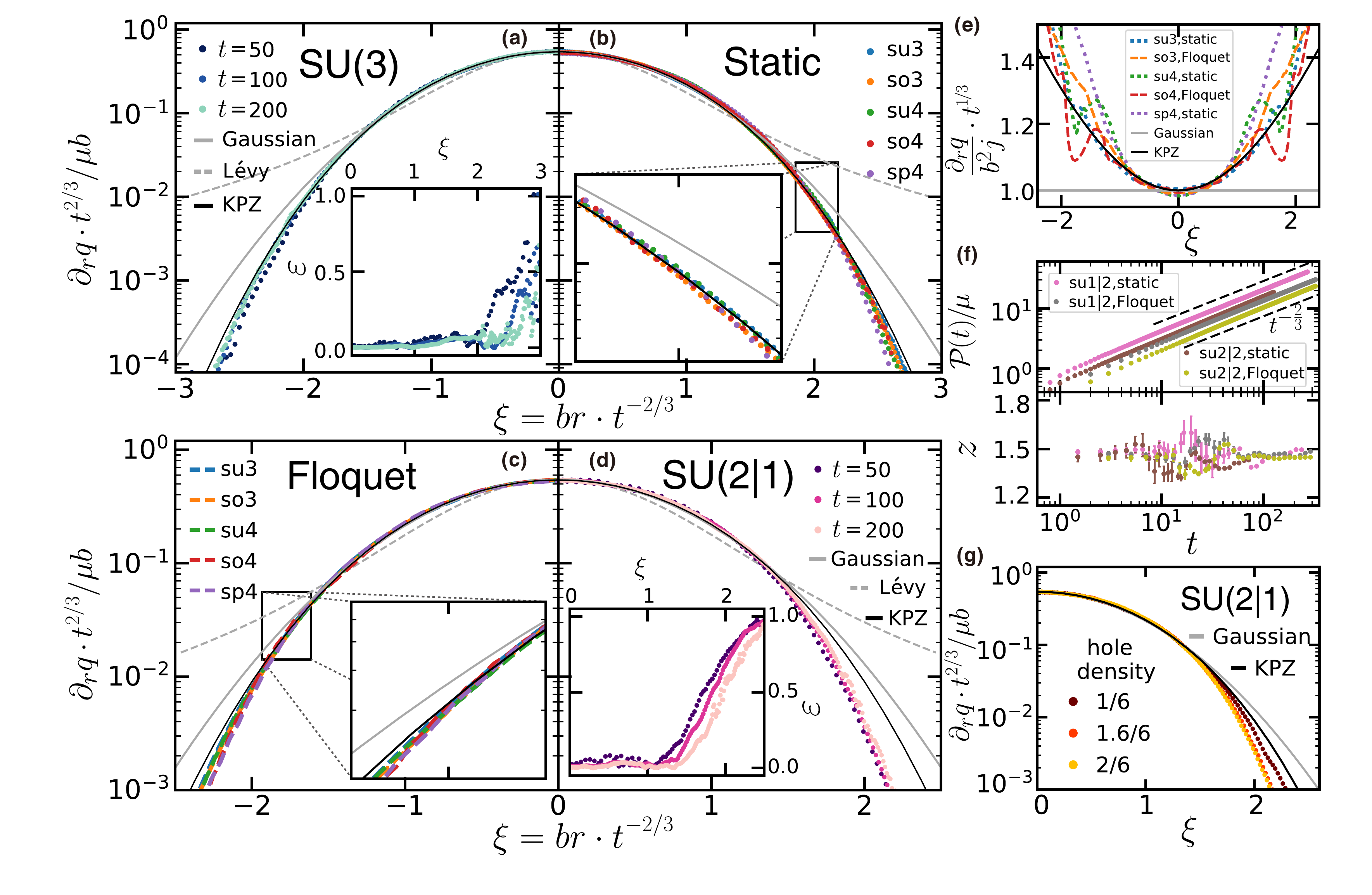}
    \caption{(a-d) The KPZ scaling function emerges from a wide variety of integrable dynamics: static, non-Abelian-symmetric models, their Floquet counterparts, and supersymmetric models. 
     (a)[(d)] At late times, the rescaled polarization profiles of the SU(3)[SU(2$|$1)] model differ from both the Gaussian and L\'{e}vy-flight expectations, but exhibit excellent agreement with the KPZ scaling function. 
     Insets of (a)[(d)]: relative difference with respect to the KPZ scaling function. We note that the agreement extends to longer length-scales as time is increased. 
     (b)[(c)] Late-time, rescaled polarization profiles of static [Floquet] integrable models with different non-Abelian symmetries.
     For all symmetries explored, the dynamics exhibit excellent agreement with the KPZ scaling function. 
     Insets of (b)[(c)]: zoom-in of the polarization profiles. (e) For all models considered, the ratio between the polarization gradient and the current is inhomogeneous, in stark contrast with the expectation for any linear transport equation.
     The observed curvature is instead in agreement with KPZ transport. 
     (f) In integrable supersymmetric models, the total charge transferred across the domain wall (upper panel) and the extracted dynamical exponent $z$ (lower panel) are consistent with superdiffusion. (g) Polarization gradients in an integrable SU(2$|$1) model with varying hole density. At the same evolution time,  systems with a smaller hole density are closer to the KPZ expectation. 
    }
    \label{fig3}
\end{figure*}
}

\begin{document}

\title{Universal Kardar-Parisi-Zhang dynamics in integrable quantum systems}

\author{Bingtian Ye}
\thanks{These authors contributed equally to this work.}
\affiliation{Department of Physics, University of California, Berkeley, California 94720 USA}
\affiliation{Department of Physics, Harvard University, Cambridge, Massachusetts 02138 USA}

\author{Francisco Machado}
\thanks{These authors contributed equally to this work.}
\affiliation{Department of Physics, University of California, Berkeley, California 94720 USA}
\affiliation{Materials Sciences Division, Lawrence Berkeley National Laboratory, Berkeley, CA 94720, USA}

\author{Jack Kemp}
\thanks{These authors contributed equally to this work.}
\affiliation{Department of Physics, University of California, Berkeley, California 94720 USA}

\author{Ross B. Hutson}
\affiliation{JILA, National Institute of Standards and Technology, Boulder, CO, 80309, USA}
\affiliation{Department of Physics, University of Colorado, Boulder, CO, 80309, USA}

\author{Norman Y. Yao}
\affiliation{Department of Physics, University of California, Berkeley, California 94720 USA}
\affiliation{Department of Physics, Harvard University, Cambridge, Massachusetts 02138 USA}
\affiliation{Materials Sciences Division, Lawrence Berkeley National Laboratory, Berkeley, CA 94720, USA}

\begin{abstract}
Although the Bethe ansatz solution of the spin-1/2 Heisenberg model dates back nearly a century, the anomalous nature of its high-temperature transport dynamics has only recently been uncovered.
Indeed, numerical and experimental observations have  demonstrated that spin transport in this paradigmatic model falls into the Kardar-Parisi-Zhang (KPZ) universality class.
This has inspired the significantly stronger conjecture that KPZ dynamics, in fact, occur in \emph{all} integrable spin chains with non-Abelian symmetry. 
Here, we provide extensive numerical evidence affirming this conjecture.
Moreover, we observe that KPZ transport is even more generic, arising in both supersymmetric  and periodically-driven models.
Motivated by recent advances in the realization of SU($N$)-symmetric spin models in alkaline-earth-based optical lattice experiments, we propose and analyze a protocol to directly investigate the KPZ scaling function in such systems. 

\end{abstract}

\maketitle

First proposed in the context of surface growth~\cite{kardar:1986}, the Kardar-Parisi-Zhang (KPZ) equation has become central to our understanding of many stochastic processes~\cite{halpin-healy:1995,corwin:2011,corwin:2014}.
While the central limit theorem ensures that the late-time physics of  linear stochastic processes is typically Gaussian, the KPZ equation evades this fate. 
Instead, it represents a distinct universality class which emerges in myriad dynamical phenomena, ranging from directed polymers and traffic models to kinetic roughening~\cite{baik1999distribution,calabrese2010free,dotsenko2010bethe,amir2011probability,popkov:2015,corwin2018operatorname,matetski2016kpz,corwin2014tropical,degier:2019,corwin2020stochastic}.

The characterization of dynamical universality classes requires one to specify both the scaling exponents and functions of the theory.
This is perhaps most familiar in the context of Brownian motion, where the diffusive late-time behavior follows a Gaussian scaling function;
the width of the corresponding distribution  grows as  $\sim t^{1/z}$, where $z=2$ is the  dynamical scaling exponent.
By contrast, the scaling functions for the KPZ universality class are significantly more complex and their exact functional form represents a relatively recent mathematical achievement~\cite{bertini1995stochastic,prahofer:2004,hairer2013solving,amir2011probability,sasamoto2010exact,corwin2018exactly}.
The associated dynamical scaling exponent is neither diffusive nor ballistic ($z=1$), but rather superdiffusive with $z=3/2$. %

Typically, KPZ behavior is expected in non-linear, out-of-equilibrium \emph{classical} systems subject to external noise; in this context, its observation is extremely robust and  does not require any fine-tuning or the presence of a particular symmetry.
To this end, the numerical and experimental observation of KPZ universality in a one-dimensional \emph{quantum} spin-chain (i.e.~the spin-$1/2$ Heisenberg model), fine-tuned for \emph{both} integrability and SU(2) symmetry, has attracted widespread attention~\cite{ljubotina:2017,ilievski:2018,gopalakrishnan:2019,ljubotina:2019,rubio-abadal:2019,ilievski2021superuniversality,wei:2021,scheie2021detection}.
Interestingly, this observation is at odds with  conventional expectations for spin chain transport, which  predict diffusion~\cite{jeon:1996,abanov:2006,bettelheim:2006,blake:2018}.
This naturally motivates the following question: Is the Heisenberg chain an isolated exception, or the first example of a broader group of quantum models in the KPZ universality class?

Seminal recent work has made elegant progress on this question by proving that \emph{all} integrable spin chains with a non-Abelian symmetry exhibit superdiffusive transport with $z=3/2$ (Fig.~\ref{fig1})~\cite{ilievski2021superuniversality}. 
However, a single scaling exponent does not uniquely specify the universality class and no analysis has been able to determine the nature of the corresponding scaling functions.

\figOne 

In this Letter, we present an extensive numerical investigation that supports the following stronger conjecture --- the dynamics of all integrable spin chains with a non-Abelian symmetry belong to the KPZ universality class.
Leveraging a novel tensor-network-based technique dubbed density matrix truncation (DMT) \cite{white:2018, ye:2020}, we demonstrate that the spin dynamics of such models are precisely \emph{captured by the KPZ scaling function} (Fig.~\ref{fig3}).
Intriguingly, our numerical observations suggest that the conjecture holds  not only for static systems, but also for periodically driven (Floquet) systems~\cite{vanicat:2018,ljubotina:2019}, as well as supersymmetric models.

By applying perturbations to break either the non-Abelian symmetry or the integrability, we characterize the approach to superdiffusive transport from regimes where there is analytical control on the dynamics. 
We reproduce these analytical results with unprecedented accuracy, both verifying and benchmarking our numerics, as well as providing independent evidence for the purported microscopic mechanism underlying  superdiffusion~\cite{gopalakrishnan:2019,De_Nardis_2019}.
Finally, we propose an experimental implementation --- based upon alkaline-earth atoms in optical lattices --- capable of investigating KPZ transport in a variety of SU($N$)-symmetric, integrable models.

In this work, we study the universality classes describing the infinite-temperature dynamics for a variety of one-dimensional quantum spin-chains. 
We will focus on the dynamics of a locally conserved charge $\hat{Q} = \sum_{r} \hat{q}_r$, typically spin. 
If the system is characterized by a dynamical universality class,  at late times the correlation function must collapse under an appropriate rescaling of space and time:
\begin{equation}
\langle\hat{q}_r(t)\hat{q}_0(0)\rangle_{T=\infty} \propto t^{-1/z} f\left(\frac{r}{t^{1/z}}\right).
\end{equation}
This collapse defines the dynamical scaling exponent $z$ and the scaling function $f(\xi)$, which together determine the universality class.

\emph{Probing transport dynamics.}---Let us  begin by exploring the dynamical exponent. 
While $z$ can in principle be extracted from the behavior of  $\langle\hat{q}_r(t)\hat{q}_0(0)\rangle_{T=\infty}$, a simpler and more robust numerical setup is to consider the dynamics of a domain wall. 
More specifically, we perturb an infinite-temperature density matrix  with a weak domain-wall-like imbalance in the charge density (Fig.~\ref{fig1}a):
\begin{equation}
    \rho(t=0) \propto (\mathds{1}+\mu \hat{q})^{\otimes L/2} \otimes (\mathds{1}-\mu \hat{q})^{\otimes L/2}, \label{eq:denmat}
\end{equation}
where $\mu$ determines the strength of the perturbation and $L$ is the length of the chain.

As the system equilibrates,  charge crosses the domain wall---%
the precise details of how this occurs reveals  properties of the dynamical universality class [Fig.~\ref{fig1}(b)].
In particular, we focus on the spatial profile of the charge density  $q(r,t) = \langle \hat{q}_r(t) \rangle$ (hereafter, denoted as polarization), as a function of time $t$ and displacement $r$ from the domain wall.
A natural measure of transport is the total polarization transferred across the domain-wall, $\mathcal{P}(t) = \sum_{r=1}^{L/2} \left(\mu - q(r,t)\right)$, which provides a robust way to determine $z$: $\mathcal{P} (t) \propto t^{1/z}$.

To implement the domain-wall dynamics,  we represent $\rho$ using a matrix product density operator and compute its evolution via DMT~\cite{white:2018, ye:2020}. 
The truncation procedure in DMT is specifically designed to preserve local operators, such as the energy density, polarization, and their currents; this choice makes DMT particularly amenable for probing the universality class of many-body transport dynamics~\cite{SM}. 

Although we will explore a wide variety of integrable models (Fig.~\ref{fig3}), let us begin by focusing our discussions on the SU(3)-symmetric, spin-1 chain~\cite{uimin1970one,lai1974lattice,sutherland:1975}: \begin{equation} \label{eq:SU(3)}
  H_{\mathrm{SU(}3\mathrm{)}} = \sum_i \vec{S}_i\cdot \vec{S}_{i+1} + (\vec{S}_i \cdot \vec{S}_{i+1})^2,
\end{equation} where $\vec{S}_i$ is the vector of spin-1 operators acting on site $i$.
Figure~\ref{fig1}(b) depicts the melting of the domain wall as a function of time, starting from the initial state, $\rho(t=0)$ with $\hat{q} = \hat{S}^z$ [Eqn.~\ref{eq:denmat}].
The corresponding polarization transfer, $\mathcal{P} (t)$, exhibits a power-law $\sim t^{2/3}$ (blue line, Fig.~\ref{fig2}b), consistent with the expected $z=3/2$ exponent~\cite{dupont:2020}.
This exponent can be independently confirmed via a scaling collapse of the polarization profile (Fig.~\ref{fig1}c).

\figTwo

\figThree

In order to tune the system away from superdiffusion, one can perturb the spin-chain by either breaking the symmetry of the initial state or the symmetry of the Hamiltonian.
To study the former, we initialize the system in $\rho(t=0)$ and add a uniform magnetization, $\delta$ (along the $\hat{z}$-axis) on each site.
The polarization transfer exhibits markedly distinct dynamics with a ballistic exponent, $z=1$ (orange line, Fig.~\ref{fig2}b).
Analytically, for weak magnetizations, the velocity of this ballistic transport is expected to scale linearly with  $\delta$; this is indeed borne out by the data (Fig.~\ref{fig2}c)~\cite{gopalakrishnan:2019,denardis:2020}. 
For the spin-$1/2$ Heisenberg model, an even stronger statement can be made---the velocity extracted from DMT quantitatively agrees with analytic calculations [via generalized hydrodynamics (GHD)] even in the non-linear regime (inset, Fig.~\ref{fig2}c) \cite{gopalakrishnan:2019,De_Nardis_2019}.

Next, we break the symmetry of $H_{\mathrm{SU(}3\mathrm{)}}$  down to U(1) by considering the so-called Izergin-Korepin family of integrable spin-1 models~\cite{izergin:1984,vichirko1983excitation,hao2014exact,izergin1981inverse}. 
We parametrize the symmetry-breaking strength by $\Delta$, such that when $\Delta=0$, we recover $H_{\mathrm{SU(}3\mathrm{)}}$.
For finite values of $\Delta$, 
 we observe diffusive transport with the polarization transfer scaling as $\mathcal{P} (t) \sim t^{1/2}$ (purple line, Fig.~\ref{fig2}b).
 In addition, the extracted diffusion coefficient, $D$, diverges as $\Delta\rightarrow 0$, consistent with the approach to superdiffusion  (Fig.~\ref{fig2}d).
 The analogous numerical experiment in the Heisenberg model (where $\Delta$ controls the XXZ anisotropy) again quantitatively agrees with analytic calculations.

A few remarks are in order.
First, the agreement between DMT numerics and GHD analytics (which have different underlying assumptions) serves a dual benchmarking role; in particular, it highlights DMT's ability to faithfully characterize late-time transport dynamics and GHD's ability to quantitatively compute transport coefficients in integrable models~\cite{karrasch2014real,De_Nardis_2019}. 
Second, in addition to breaking the non-Abelian symmetry of the Hamiltonian, one can also probe the effect of integrability breaking.
To this end, we perturb $H_{\mathrm{SU(}3\mathrm{)}}$ using SU($3$)-symmetry-respecting, but \emph{integrability-breaking} next-nearest-neighbor interactions.
As expected for generic non-integrable models, $\mathcal{P} (t) \sim t^{1/2}$, consistent with diffusive transport (green line, Fig.~\ref{fig2}b)~\cite{spohn2012large,doyon2019diffusion,friedman2020diffusive}.

\emph{Probing KPZ dynamics.}---%
While our numerical observation
of a $z=3/2$ exponent in $H_{\mathrm{SU(}3\mathrm{)}}$ clearly establishes the presence of superdiffusion, it does not determine the system's dynamical universality class.
Indeed, such an exponent can also arise in long-range interacting systems exhibiting L\'evy flights, as well as rescaled diffusion~\cite{ljubotina:2017,ljubotina:2019,rubio-abadal:2019,levy1954theorie,shlesinger1995levy,joshi:2021}. 

To this end, we now investigate the universal scaling function.
In particular, using our domain-wall dynamics, we can compute
the charge correlation function from the spatial gradient of the polarization profile~\cite{ljubotina:2019}:
\begin{equation}
  \langle \hat{q}_r(t)\hat{q}_0(0)\rangle_{T=\infty}= \lim_{\mu\to 0} \frac{\partial_r q(r,t)}{2\mu} = \frac{b}{t^{2/3}} f\left(\frac{br}{t^{2/3}}\right),
\end{equation}
where $b$ is a system-dependent parameter~\footnote{We use $\partial_r$ as a short-hand for discrete difference in the our system: $\partial_r q(r,t) = \langle \hat{q}_{r+1}(t)\rangle - \langle \hat{q}_r(t) \rangle$.}.

As depicted in Figure~\ref{fig3}a, $\partial_r q(r,t)$ indeed collapses under the rescaling, $f(\xi=brt^{-2/3})$. 
For L\'{e}vy flights, one expects power-law tails (gray dashed line), which are manifestly inconsistent with the data. 
However, the difference between rescaled diffusion and KPZ is more subtle: for the former, $f(\xi)$ is Gaussian, while for KPZ, $f(\xi)$ exhibits faster decaying tails $\sim\exp\left(-0.295|\xi|^3\right)$~\cite{prahofer:2004,sasamoto2010exact,hairer2013solving}.
The data quantitatively agree with the KPZ prediction: The longer the evolution time, the closer $\partial_r q(r,t)$ is to the KPZ scaling function (highlighted by the relative error, Fig.~\ref{fig3}a inset).
This agreement allows us to directly extract $b = 0.460\pm 0.001$, which reflects the ratio between the diffusive smoothing, and the non-linear growth and noise in the KPZ equation.
We emphasize that these observations apply to \emph{any} conserved charges generated by the non-Abelian symmetry~\cite{SM}.

A complementary way to distinguish between rescaled diffusion and KPZ dynamics is to study the ratio between the spin
current, $j(r,t)= - \int_{-\infty}^{r} \partial_t q(r',t) dr'$, and the polarization gradient.
In rescaled diffusion, Fick's law  ensures that the two are proportional, $j(r,t) \propto t^{1/3} \partial_r q(r,t)$, while the non-linearity of KPZ transport  leads to the breakdown of this proportionality~\cite{prahofer:2004,ljubotina:2019}. 
Crucially, as illustrated in Fig.~\ref{fig3}e, we find that the ratio is not constant (as would be predicted for rescaled diffusion) and rather, is in good agreement with the KPZ prediction.

\emph{Universality of KPZ dynamics.}---%
We now turn our attention to the conjecture that KPZ dynamics emerge in several broad classes of integrable models.
We will focus on three distinct settings: (i) static models with generic non-Abelian symmetries, (ii) periodically-driven (Floquet) models with non-Abelian symmetries, and (iii) supersymmetric models.
In these latter two classes, even for the dynamical exponent, there are no generic results, although some particular instances are known to exhibit superdiffusion \cite{ilievski2018superdiffusion,ljubotina:2019,dupont:2020}.

The construction of static, non-Abelian, integrable spin chains has a rich history, with different prescriptions for each of the four classes of simple Lie groups: SU($N$), SO($2N$), SO($2N+1$) and SP($2N$)~\cite{korepin:1993, kassel:1995, ilievski2021superuniversality,kulish1982solutions,SM}.
As detailed in the supplementary material, we construct nearest-neighbor models with the following four symmetries, SU(4), SO(3), SO(4) and SP(4).
Following our previous strategy for $H_{\mathrm{SU(}3\mathrm{)}}$, we analyze the transport dynamics of conserved charges for each of these models. 
In all cases, we observe excellent agreement with the KPZ universality class (Fig.~\ref{fig3}b,e).

Extending this exploration to periodically driven systems requires systematically building the corresponding Floquet integrable models.
Somewhat astonishingly, one can straightforwardly build such models from their static counterparts \cite{vanicat:2018,ljubotina:2019a}.
The Hamiltonian is divided into  terms acting on even and odd bonds (denoted as $H_{\mathrm{even}}$ and $H_{\mathrm{odd}}$, respectively), which are then alternatingly applied, leading to a Floquet unitary:  $U = e^{-iH_{\mathrm{odd}}T/2} e^{-iH_{\mathrm{even}}T/2}$.
Using this procedure, we can extend our  analysis to the Floquet regime for all of the previous non-Abelian models (Fig.~\ref{fig3}c,e).
Our conclusions are identical.
The resulting transport falls within the KPZ universality class even though energy is no longer conserved.

Finally, let us consider integrable models where the non-Abelian symmetry is replaced with supersymmetry.
Such models have been conjectured to exhibit superdiffusion, but observing this, either numerically or analytically,  remains an open challenge~\cite{ilievski:2018, ilievski2021superuniversality}.
Here, we focus on a pair of spinful fermionic lattice models: the $t$-$J$ model (with $t=2J$), and the Essler-Korepin-Schoutens (EKS) model~\cite{sarkar:1991, essler:1992}. 
These exhibit the two simplest supersymmetries, SU($1|2$) and SU($2|2$), respectively.

The defining feature of models with supersymmetry is that their conserved charges fall into two types: bosonic and fermionic,  although only the bosonic charge can in principle exhibit superdiffusion~\cite{ilievski:2018}.
For the $t$-$J$ model, each lattice site can be occupied by either a spin-up fermion, a spin-down fermion, or a hole. 
The conserved bosonic charges are given by the total number of holes, and the total spin.
Holes live in the Abelian U(1) sector and thus lack particle-hole symmetry leading to a finite Drude weight and ballistic transport \cite{ilievski:2018}. 
Therefore, we study the spin polarization, given by the difference between the number of spin-up and spin-down particles. 
As before, we prepare a weak domain-wall in the spin polarization while keeping the other charge densities---including the hole density---constant.

For both the static and Floquet $t$-$J$ models, we observe superdiffusive spin transport (with $z=3/2$) via both the polarization transfer (Fig.~\ref{fig3}f) and the collapse of the polarization profile~\cite{SM}.
The numerical evidence that  spin transport falls within the KPZ universality class is more subtle.
In particular, the polarization gradient, $\partial_r q(r,t)$, exhibits a discrepancy with both the KPZ and  Gaussian expectations (Fig.~\ref{fig3}d).
However, the finite-time flow of $\partial_r q(r,t)$ approaches the KPZ scaling function in  the same qualitative fashion as is observed in the SU(3) case (insets, Fig.~\ref{fig3}a,d).
Moreover, a careful comparison of the relative error to the Gaussian model suggests that rescaled diffusion cannot be the correct limiting behavior~\cite{SM}. 

A few remarks are in order. 
First, we conjecture that  finite-time effects are exacerbated in  supersymmetric models owing to the presence of additional ballistic modes.
To test this conjecture, we decrease the hole density~\footnote{We note that when there are no holes, one recovers an SU(2)-symmetric model.}, and indeed observe an improved  convergence to KPZ universality (Fig.~\ref{fig3}g).
Curiously, this suggests that KPZ dynamics might arise in supersymmetric systems for generic fermionic filling fractions.
Second, our analysis for the EKS model arrives at identical conclusions (Fig.~\ref{fig3}f)~\cite{SM}.

\emph{Experimental proposal.}---Recent advances in the control and manipulation of alkaline-earth atoms in optical lattices have opened the door to studying SU($N$)-symmetric spin models~\cite{weitenberg:2011,cappellini:2014,pagano:2014,scazza:2014,zhang:2014,miranda2015site,yamamoto2016ytterbium,miranda2017site,sonderhouse:2020,schine2021long}.
In particular, at  unit-filling in the Mott insulating phase, the lack of hyperfine coupling in the $ns^2~{}^1S_0$ electronic ground state naturally leads to SU($N$)-symmetric spin-exchange interactions~\cite{gorshkov:2010,hofrichter:2016,ozawa:2018,sonderhouse:2020}:
\begin{equation}
    H_{\mathrm{SU(}N\mathrm{)}} = J_{\mathrm{SU(}N\mathrm{)}}\sum_{i} \sum_{\alpha,\beta=1}^N s^{\alpha,\beta}_{i}s^{\beta,\alpha}_{i+1}, 
\label{eq:SUN}
\end{equation}
where $s^{\alpha,\beta}_{i} = \ket{\alpha}\bra{\beta}$ on site $i$;  in one dimension, $H_{\mathrm{SU(}N\mathrm{)}}$ is integrable and precisely corresponds to the models considered above (e.g.~Eqn.~\ref{eq:SU(3)}).

The observation and characterization of KPZ transport requires the ability to address two main experimental challenges: (i) preparing  near infinite-temperature states with a well-defined domain-wall polarization and (ii) measuring the tails of the scaling function with sub-percent accuracy. 
The former can be accomplished via a two step process: 
first, optical pumping via an intercombination transition (e.g.~$ns^2~{}^1S_0 \leftrightarrow nsnp~{}^3P_1$) can be used to generate arbitrary magnetization distributions which are preserved upon  cooling to the Mott insulator~\cite{SM}; second, with single-site addressing~\cite{bakr:2010,sherson:2010,cheuk:2015, parsons:2015, omran:2015, haller:2015, wei:2021}, a coherent optical drive can be applied to half the system in order to prepare the domain wall.

Achieving the latter is significantly more subtle. 
In order to distinguish between KPZ dynamics and rescaled diffusion, careful estimates suggest the need to experimentally resolve the scaling function with a relative error of $\sim 10^{-3}$~\cite{SM}.
Achieving this error floor requires the ability to spatially resolve spin-transport dynamics over long time-scales and large distances. 
For concreteness, let us consider $^{87}$Sr atoms loaded into a two-dimensional optical lattice~\cite{sonderhouse:2020, park:2021,bothwell_resolving_2022}.
Recent experiments have demonstrated the elegant use of cavity-enhancement to realize homogeneous lattices capable of supporting Mott insulators with a diameter of $\sim 300$ sites~\cite{park:2021, SM}.
By implementing strong confinement in one direction, one can subsequently divide the system into $\sim 250$ independent chains, each with length $\sim 150$ sites.
Assuming an on-site interaction energy, $U \sim 3~\mathrm{kHz}$, and a tunneling rate,  $t \sim 300~\mathrm{Hz}$, yields a spin-exchange interaction, $J = 2t^2/U \approx 60~\mathrm{Hz}$~\cite{park:2021, SM}.
Optimizing for an evolution time of $\sim 50/J$ and assuming an experimental cycle time of $\sim 10~\mathrm{s}$~\cite{sonderhouse:2020}, we estimate that a relative error of $\sim 10^{-3}$, can be achieved within two days of averaging~\cite{SM}. 
Finally, the presence of a finite density ($\gtrsim$ 1\%~\cite{SM, ibarra-garcia-padilla:2021}) of doublons and holes in the Mott insulator will perturb the polarization dynamics, but the exact nature of their effect remains an intriguing open question.

\emph{Acknowledgements}---We gratefully acknowledge discussions with M. Aidelsburger, I. Bloch, V. Bulchandani, A. Kaufman,  J. Moore, and J. Ye.
We are especially indebted to S. Gopalakrishnan and R. Vasseur for introducing us to the framework of generalized hydrodynamics and for collaborations on related works. 
This work was supported by the  U.S. Department of Energy, Office of Science, National Quantum Information Science Research
Centers, Quantum Systems Accelerator (QSA), by the AFOSR MURI program (FA9550-21-1-0069), by the David and Lucile Packard foundation, and by the Alfred P. Sloan foundation. 
F.M. is supported by the Army Research Office (grant no. W911NF2110262).
J. K. is supported by the U.S.
Department of Energy through the Quantum Information
Science Enabled Discovery (QuantISED) for High Energy
Physics (KA2401032) and through the GeoFlow Grant
No. de-sc0019380.

\bibliographystyle{apsrev4-1}
\bibliography{KPZ} 

\end{document}


\title{Supplementary Information: \\
Universal Kardar-Parisi-Zhang dynamics in integrable quantum systems}

\author{Bingtian Ye}
\thanks{These authors contributed equally to this work.}
\affiliation{Department of Physics, University of California, Berkeley, California 94720 USA}
\affiliation{Department of Physics, Harvard University, Cambridge, Massachusetts 02138 USA}

\author{Francisco Machado}
\thanks{These authors contributed equally to this work.}
\affiliation{Department of Physics, University of California, Berkeley, California 94720 USA}
\affiliation{Materials Sciences Division, Lawrence Berkeley National Laboratory, Berkeley, CA 94720, USA}

\author{Jack Kemp}
\thanks{These authors contributed equally to this work.}
\affiliation{Department of Physics, University of California, Berkeley, California 94720 USA}

\author{Ross B. Hutson}
\affiliation{JILA, National Institute of Standards and Technology, Boulder, CO, 80309, USA}
\affiliation{Department of Physics, University of Colorado, Boulder, CO, 80309, USA}

\author{Norman Y. Yao}
\affiliation{Department of Physics, University of California, Berkeley, California 94720 USA}
\affiliation{Department of Physics, Harvard University, Cambridge, Massachusetts 02138 USA}
\affiliation{Materials Sciences Division, Lawrence Berkeley National Laboratory, Berkeley, CA 94720, USA}

\maketitle
\section{Convergence of the DMT method}
\begin{figure}[h!]
\includegraphics[width=5.5in]{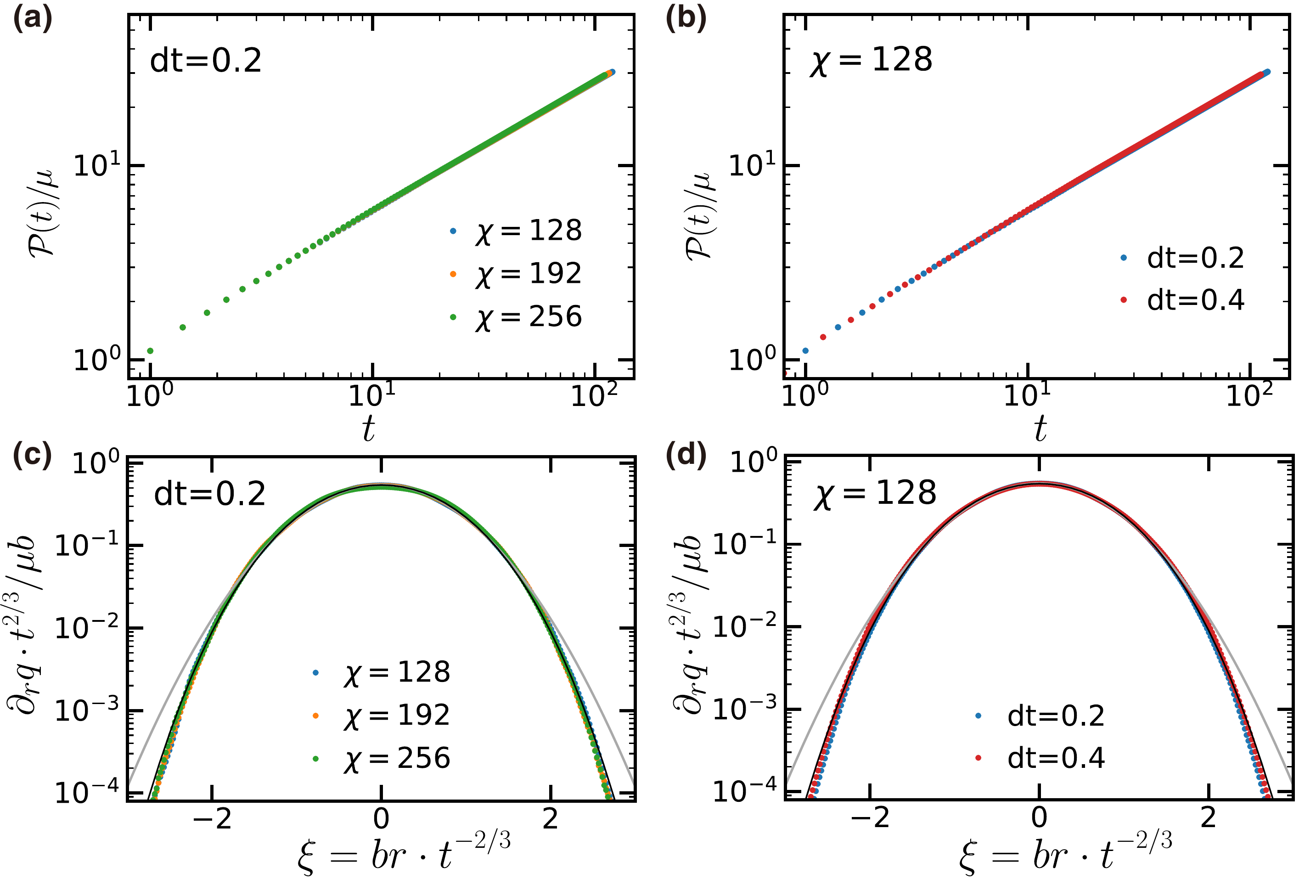}
\centering
\caption{
(a)(b) Convergence of total population transfer as a function of the bond dimension $\chi$ and the Trotter step length $dt$. 
(c)(d) Convergence of population profile of the bond dimension $\chi$ and the Trotter step length $dt$. 
}
\label{fig:PConv}
\end{figure}

Similar to other TEBD-like numerical methods, the accuracy of the DMT method is controlled by the Trotter step size $dt$ and the bond dimension $\chi$. 
In this section, we show the convergence of our numerical results with respect to them. 
In particular, we simulate the same dynamics with $\chi\in \{128,192,256\}$, and $dt \in \{0.4,0.2\}$. 
For both the total population transfer and the population profile, we observe good agreement across different tuning parameters (Fig.~\ref{fig:PConv}). 
This confirms that the results presented in the main text are of high enough precision to distinguish between the KPZ and the (rescaled) Gaussian expectations. 

\section{Finite-time effects and finite-size effects in KPZ dynamics}

Although the precise observation of KPZ scaling is restricted to the infinitely late-time behavior of an infinitely long chain, any numerical investigation will always be limited to the finite time dynamics of a finite system.
Given such limitations, we perform a detailed analysis of how finite-size effects impact our observation of KPZ dynamics to ensure that they do not affect our conclusions.

\subsection{Finite-time flow of the polarization profile}
\begin{figure}[h!]
\includegraphics[width=5.5in]{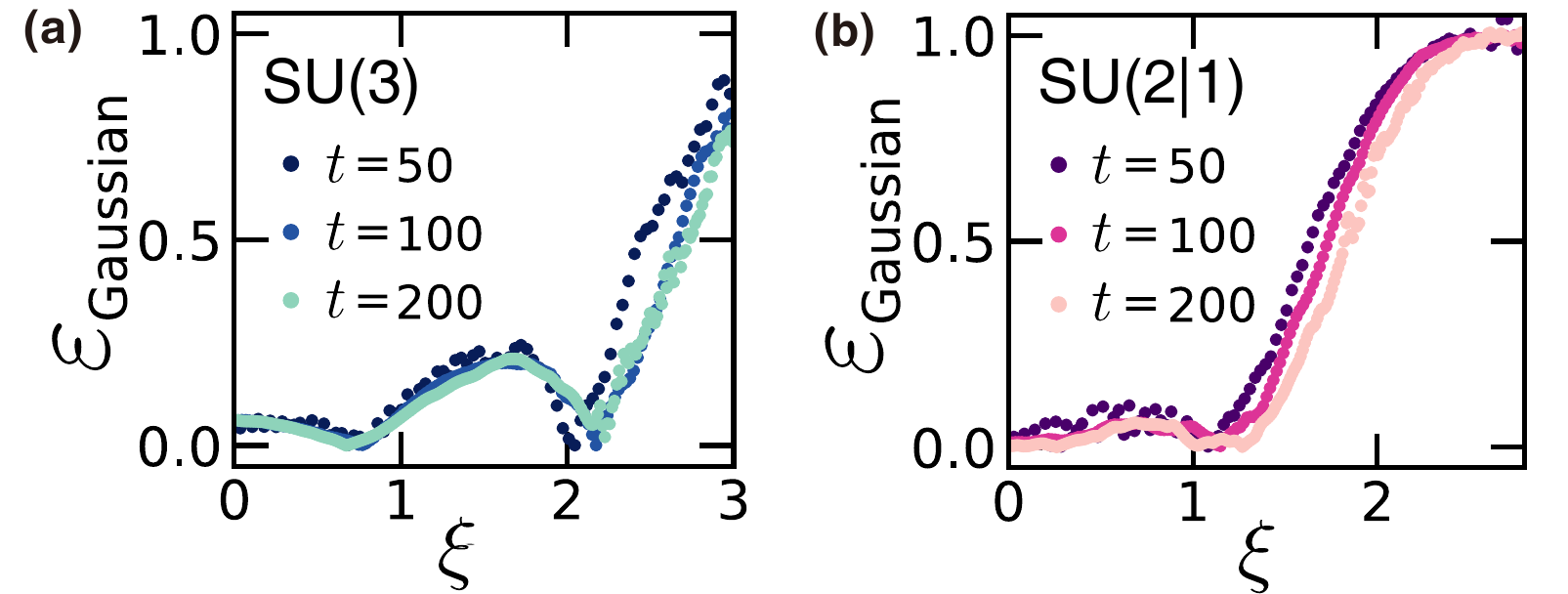}
\centering
\caption{Relative error between the population density and the Gaussian expectations for
(a) the SU(3) model and (b) the SU(1$|$2) model. 
}
\label{fig:TConv}
\end{figure}

At any finite time, the polarization profile will not exactly match the KPZ expectation; 
however, this agreement improves at later times, as highlighted in Fig.~3(a) inset and Fig.~3(d) inset of the main text using the relative difference between the numerical results and the KPZ scaling function. 
In order to confirm that the observed dynamics are not governed by a different transport equations, e.g., (rescaled) Gaussian transport, we measure the relative error at different times between the numerics and the Gaussian profile, in addition to the direct comparison against the Gaussian expectation shown in Fig.~2(c) of the main text. 
Crucially, we observe that the errors do \emph{not} decrease to zero as time goes on [SU(3) model in Fig.~\ref{fig:TConv}(a) and SU(1$|$2) model in Fig.~\ref{fig:TConv}(b)]. 

\subsection{Ballistic modes and the corresponding finite-size effect}
To understand the finite-size effect in our simulations, recall that, although the charge transport is diffusive, the underlying quasiparticles still propagate ballistically. 
These ballistic modes are responsible for the ballistic transport of energy, and are more sensitive to finite-size effects: they move across the system parametrically faster than the superdiffusive modes. 
For the domain wall initial state, the energy density is inhomogeneous with a peak at the domain wall position. 
In the subsequent evolution, the quasiparticles that carry energy will move ballistically across the chain, bounce back from the boundary of the chain and eventually affect the superdiffusive charge dynamics. 
This happens much faster that it takes for the smoothing out of charge domain wall to reach the boundary. 
In this subsection, we demonstrate this effect in the prototypical SU(3) model. 

\begin{figure}[h!]
\includegraphics[width=6in]{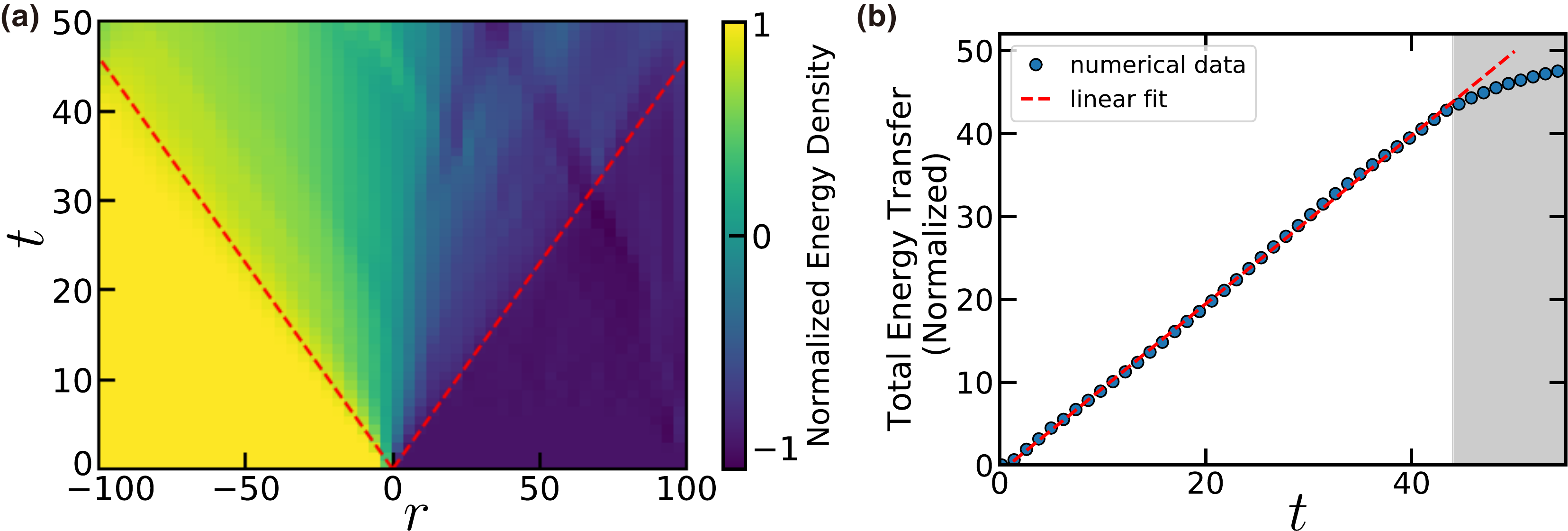}
\centering
\caption{Ballistic transport of energy in the SU(3) integrable model. (a) Spatial-temporal profile of local energy density. The red dashed line marks the light-cone of the ballistic transport of energy. (b) Total energy transferred across the initial domain wall grows linearly as a function of time. In the shaded region, the data deviates from linear growth due to the finite-size effect, perfectly consistent with the line-cone speed extracted from (a). 
}
\label{fig:energytrans}
\end{figure}

To estimate the timescale for the ballistic modes to affect the charge dynamics, we study the velocity of the energy transport. 
In particular, analogous to the setup where we study the charge transport, we initialize the system with a domain-wall-like energy density, and consider how it smooths out. 
In Fig.~\ref{fig:energytrans}, we observe a clean light cone of the energy density, from which we extract a velocity of $\sim 2.2$ (in units of lattice constant per $1/J$). 
We can also look at the total energy transferred across the domain wall. 
Unlike the charge transport, we observed a linear growth of energy, and can extract the transport velocity from the growth rate, consistent with the one measured from the light cone. 

\begin{figure}[h!]
\includegraphics[width=3in]{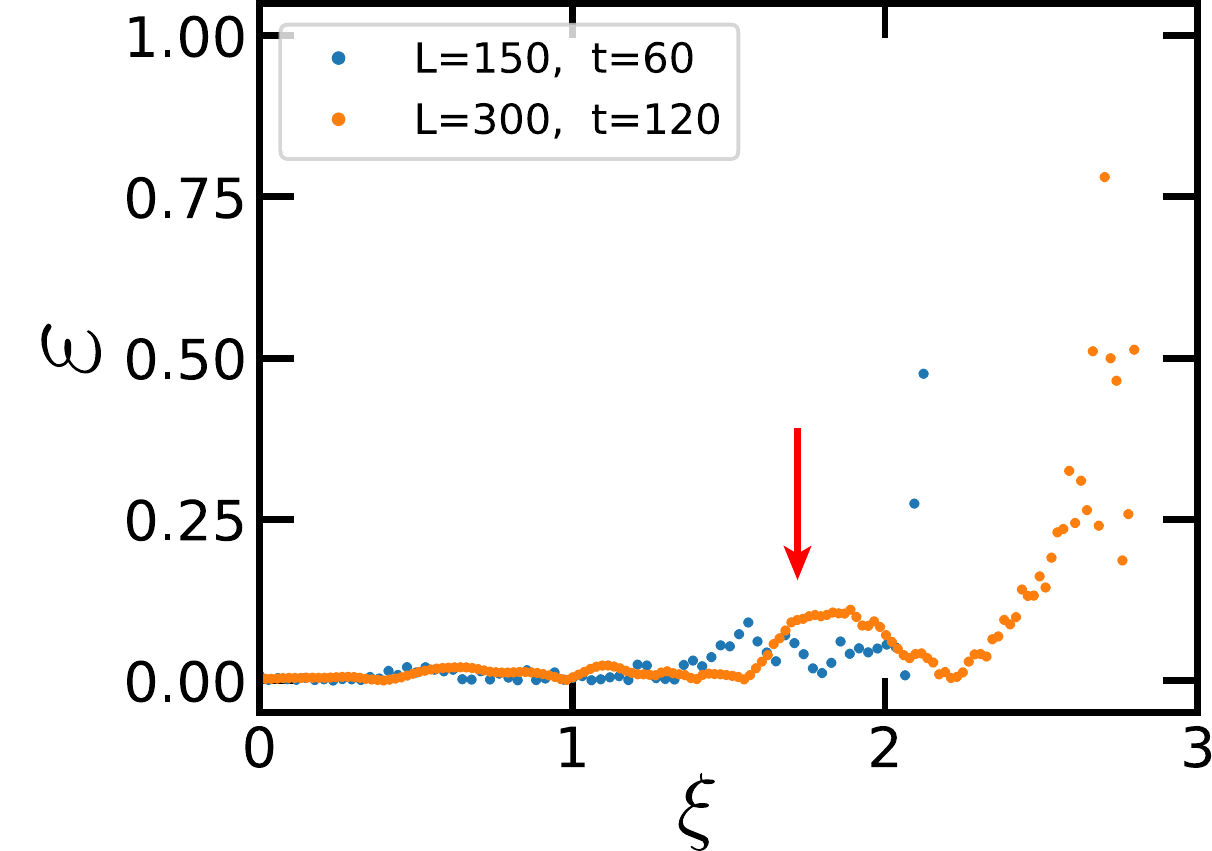}
\centering
\caption{Relative difference between numerics and the KPZ scaling function. The red arrow marks the deviation caused by the finite-size effect, whose timescale is consistent with the light-cone velocity in the system. As the system size doubled (for $L=150$ to $L=300$), the time for the finite-size to appear also gets roughly doubled. 
}
\label{fig:ballistic_err}
\end{figure}

Based on this velocity, we can estimate the time for the ballistic mode to reach the boundary as $\sim 135$ for a system size $L=600$. 
In particular, for an initial domain wall of polarization, the energy density is homogeneous everywhere except at the domain wall. 
As discussed above, the finite-size effect which limits the polarization transport occurs when the ballistic component reflects from the boundary of the system. 
Given a system size of $L=600$, the time for finite-size effects to affect the charge density dynamics can then be estimated to be $135 < t\lesssim 270$. 
Therefore, we choose $t=100,200$ when comparing our numerical results against KPZ scaling functions (Fig.~3a in the main text). 
In systems with smaller sizes, we can easily observe such finite-size effects in the relative error between the numerics and the KPZ expectation. 
As highlighted in Fig.~\ref{fig:ballistic_err}, this finite-size effects results in a ``bump" in the relative error resulting from the ballistic transport that bounces back from the boundary. 
Indeed, for system sizes of $L=150$ and $L=300$, the bump emerges at around $t=60$ and $t=120$ respectively, in agreement with our estimate of the ballistic velocity.

\section{Details of the models}
In this section, we explicitly list the matrix form of all the Hamiltonians we studied in the main text, including those with each of the four classes of simple non-Abelian symmetries, those with super-symmetries, the symmetry-breaking model, and the integrability-breaking model. 

To simplify the notation, we use $S^{\alpha,\beta}$ to denote the matrix with all elements being 0 except the $\alpha^{th}$ row and the $\beta^{th}$ column which takes value 1: $S^{\alpha,\beta} = |\alpha\rangle \langle \beta|$.
It is easy to notice
\begin{equation}
[S^{\alpha,\beta}, S^{\mu,\nu}] = \delta_{\beta,\mu} S^{\alpha,\nu}-\delta_{\nu,\alpha}S^{\mu,\beta}. 
\label{eq:commutation}
\end{equation}
We note that one can use any sets of operators satisfying this commutation relation to replace $S^{\alpha,\beta}$ in the following Hamiltonians, and also obtain integrable models with the same symmetries. 

\subsection{Non-Abelian symmetry}
In the main text, we consider models with the non-Abelian symmetries that can be described by simple Lie algebras. 
The simple Lie algebras fall into four classical series $A_N$ [SU($N$)], $B_N$ [SO($2N+1$)], $C_N$ [SP($2N$)] and $D_N$ [SO($2N$)], and five exceptional cases $G_2$, $F_4$, $E_6$, $E_7$, and $E_8$. 
Here, we only focus on the classical series. 

Let us define two operators acting on site $i$ and $j$ as
\begin{equation}
\begin{split}
\Pi_{ij} &= \sum_{\alpha,\beta} S^{\alpha,\beta}_i \otimes S^{\beta,\alpha}_j\\
\Xi_{ij}^{\mathrm{SO}} &= \sum_{\alpha,\beta} S^{\alpha,\beta}_i \otimes S^{N-\alpha,N-\beta}_j\\
\Xi_{ij}^{\mathrm{SP}} &= \sum_{\alpha,\beta} \theta_{\alpha}\theta_{\beta}S^{\alpha,\beta}_i \otimes S^{N-\alpha,N-\beta}_j
\end{split}
\label{eq:SymH1}
\end{equation}
where $N$ is the dimension of the local Hilbert space, and $ \theta_{\alpha}=\delta_{1\le\alpha\le \frac{N}{2}}-\delta_{\frac{N}{2}+1\le\alpha\le N}$. 
We can then write down the two-site Hamiltonian for different symmetries as
\begin{equation}
\begin{split}
H^{\mathrm{SU}}_{ij} &= \Pi_{ij}\\
H^{\mathrm{SO}}_{ij} &= \Pi_{ij}-\frac{2}{N-2}\Xi_{ij}^{\mathrm{SO}}\\
H^{\mathrm{SP}}_{ij} &= \Pi_{ij}+\frac{2}{N+2}\Xi_{ij}^{\mathrm{SP}}
\end{split}
\end{equation}
with appropriate $N$~\cite{ilievski2021superuniversality,korepin:1993,kassel:1995,kulish1982solutions}.

The integrable static model then has the Hamiltonian $H = \sum_i H^{\mathrm{sym}}_{i,i+1}$, where ``sym" can be SU, SO or SP. 
For the integrable Floquet model, we further define:
$H_{\mathrm{even}} = \sum_i H^{\mathrm{sym}}_{2i,2i+1}$ and $H_{\mathrm{odd}} = \sum_i H^{\mathrm{sym}}_{2i-1,2i}$. 
The Floquet evolution is generated by alternately evolving the system with $H_{\mathrm{even}}$ and $H_{\mathrm{odd}}$ for time $T/2$, where $T$ is the Floquet period. 

\subsection{Super-symmetry}
Similar to the fact that the non-Abelian symmetries are characterized by Lie algebras, super-symmetries are characterized by Lie super-algebra, which is a generalization of Lie algebra with a $\mathbb{Z}_2$-grading.
Moreover, the simple Lie super-algebras also fall into two classical series (i.e. superunitary and orthosymplectic) plus several exceptional cases. 
Here, we focus on the superunitary series denoted as SU($M|N$), as a generalization of the unitary series of Lie algebra with a $\mathbb{Z}_2$-grading.
A $\mathbb{Z}_2$-grading can be understood as follows: Lie algebras describe the commutation relation between the generators of symmetry groups; with $\mathbb{Z}_2$-grading, the generators are divided into two sets; while the generators from the same set are still defined according to commutation relation, those from different groups are defined by an anti-commutation relation. 
In particular with the SU($M|N$) supersymmetry, the $(M+N)$-dimensional local Hilbert space is spanned by two sets of states, each of which includes $M$ and $N$ states respectively. 
The transition operators $S^{\alpha,\beta}$ between any two states are then  classified according to whether the two states are from the same set or not. 
Based on this intuition, it is natural to expect that an SU($M|N$) Hamiltonian has a similar form of an SU(m+n) Hamiltonian with some additional minus sign for certain terms in Eq.~\ref{eq:SymH1}. 
Indeed, the two-site Hamiltonian of an integrable system with supersymmetry SU(m$|$n) can be written as:
\begin{equation}
H^{\mathrm{Sup}}_{ij} =
\Pi^{\mathrm{Sup}}_{ij} = \sum_{\alpha,\beta}(-1)^{P(\alpha)\cdot P(\beta)} S^{\alpha,\beta}_i \otimes S^{\beta,\alpha}_j,
\label{eq:SupSymH}
\end{equation}
where $P(\alpha)=0$ if $\alpha\le M$, otherwise $P(\alpha)=1$. 
Similar to the symmetric case, the static Hamiltonian is the sum of all two-site operators, while the Floquet dynamics is generarted by alternately applying two-site Hamiltonians on even and odd bonds.  

\subsection{Symmetry breaking}
To break the SU(3) symmetry while still keeping the model integrable, we considered the Izergin-Korepin model, whose Hamiltonian only consists of nearest-neighbor interaction, i.e., $H^{\mathrm{I-K}}=\sum_i H^{\mathrm{I-K}}_{i,i+1}$, where the two-body interaction is written as \cite{izergin1981inverse,vichirko1983excitation,hao2014exact}: 
\begin{equation}
\begin{split}
    H^{\mathrm{I-K}}_{i,j}& =\frac{1}{\ch\,3\Delta\;\ch\,2\Delta}\{ \cosh5\Delta\;(S^{1,1}_iS^{1,1}_j+S^{3,3}_iS^{3,3}_j)+\sh\,2\Delta\;(\sh\,3\Delta-\ch\,3\Delta)(S^{1,1}_iS^{2,2}_j+S^{2,2}_iS^{3,3}_j)\\
    &+\sh\,2\Delta\;(\sh\,3\Delta+\ch\,3\Delta)(S^{2,2}_iS^{1,1}_j+S^{3,3}_iS^{2,2}_j)+2\,\sh\,\Delta\;\sh\,2\Delta\;(e^{-2\Delta}S^{1,1}_iS^{3,3}_j+e^{2\Delta}S^{3,3}_iS^{1,1}_j)\\
    &+\ch\,\Delta\;(S^{1,3}_iS^{3,1}_j+S^{3,1}_iS^{1,3}_j)+\ch\,3\Delta\;(S^{1,2}_iS^{2,1}_j+S^{2,1}_iS^{1,2}_j+S^{2,2}_iS^{2,2}_j+S^{2,3}_iS^{3,2}_j+S^{3,2}_iS^{2,3}_j)\\
    &-e^{-2\Delta}\,\sh\,2\Delta\;(S^{1,2}_iS^{3,2}_j+S^{2,1}_iS^{2,3}_j)+e^{2\Delta}\,\sh\,2\Delta\;(S^{2,3}_iS^{2,1}_j+S^{3,2}_iS^{1,2}_j)\}. 
\end{split}
\end{equation}
It is easy to notice that the Izergin-Korepin Hamiltonian always has a U(1) conserved charge $\sum_i (S^{1,1}_i-S^{3,3}_i)$, and the SU(3) symmetry recovers when $\Delta=0$. 

\subsection{Integrability breaking}
To break the integrability of the model while keeping the symmetry, for the SU(3) model studied in the main text, we add next-nearest-neighbor interaction to the system. 
This modification is in general also valid for models with other symmetries. 
To be specific, the non-integrable symmetric Hamiltonian can be written as:
\begin{equation}
    H = \sum_{i} H_{i,i+1}+J_{nnn}H_{i,i+2},
    \label{eq:intbreaking}
\end{equation}
where $J_{nnn}$ is the strength of the next-nearest-neighbor interaction. 

The intuition for why the integrability is broken is simple. 
The integrability of the nearest-neighbor-interacting models results from a special property that any multi-body scattering process in the system is reducible to a series of two-body scattering. 
This is further guaranteed by a combination of Yang-Baxter equation and the fact that all scattering processes follow a certain order. 
To be specific, imagine three particles placed on different sites undergo a scattering process. With only nearest-neighbor interaction, the leftmost particle and the rightmost one can interact only after either of them scatters with the middle one, which sets the order of the scattering process. 
However, the next-nearest-interacting term can break such order, and thus break the reducibility of the multi-body scattering. 
\section{Extra numerical results for KPZ superdiffusion}
\subsection{Dynamics of different charges}
\begin{figure}[h!]
\includegraphics[width=5.5in]{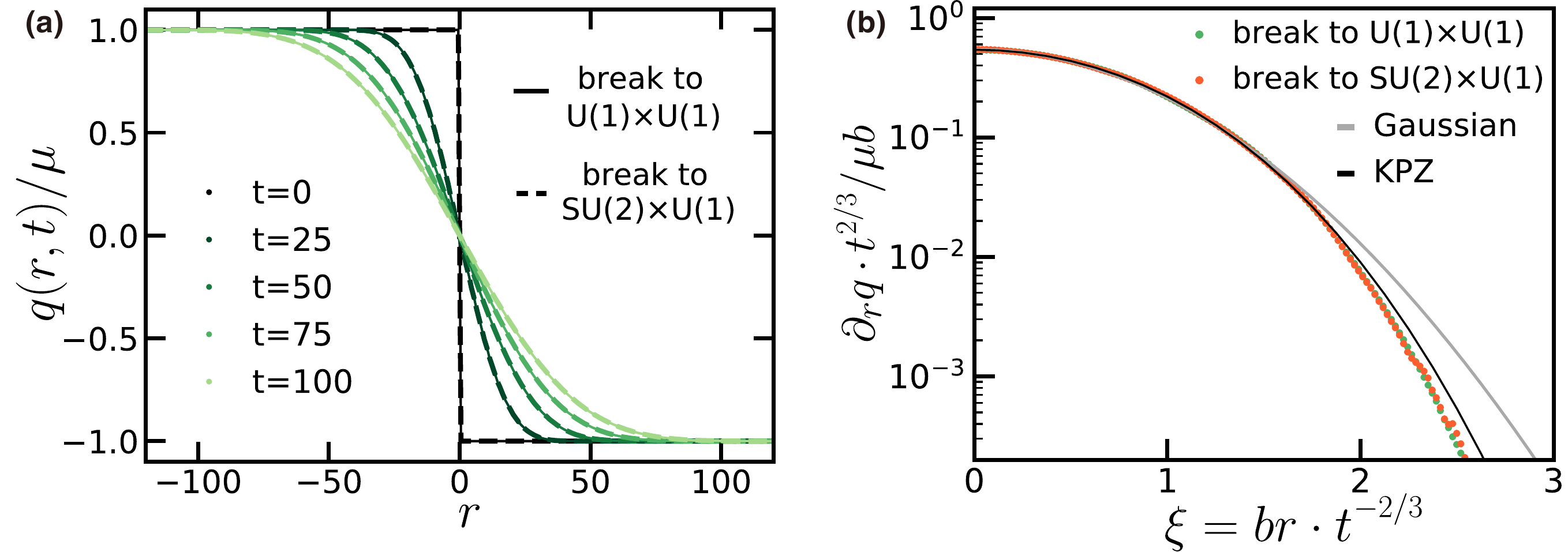}
\centering
\caption{Starting from two different domain-wall states (corresponding to the two different charges), the integrable SU(3) model exhibits the same KPZ dynamics. (a) Polarization profile. (b) Rescaled polarization gradient, equivalent to two-point correlation. The spatial profile is measured at $t=100$. 
}
\label{fig:diff_charge}
\end{figure}

In the main text, we only consider the initial state of type: 
\begin{equation}
    \rho(t=0) \propto (\mathds{1}+\mu \hat{q})^{\otimes L/2} \otimes (\mathds{1}-\mu \hat{q})^{\otimes L/2},
\end{equation}
where $\hat{q}=S^{1,1}-S^{N,N}$.
While this is the only possible initial domain-wall state in the SU(2) model (up to a global rotation), a generic non-Abelian symmetric model allows for different types of initial states corresponding to different (inequivalent) ways of perturbing the infinite temperature state. 
Here we take SU(3) model as an illustration. 
There are two inequivalent initial domain-wall states, with $\hat{q}=S^{1,1}-S^{3,3}$ and $\hat{q}=S^{1,1}+S^{2,2}-2S^{3,3}$, which breaks the symmetry of the state to U(1)$\times$U(1) and SU(2)$\times$U(1), respectively. 
Interestingly, the subsequent dynamics are in agreement. Namely, they both fall into the superdiffusive KPZ universality and have with the same superdiffusion coefficient (Fig.~\ref{fig:diff_charge}). 
We remark that understanding the robustness of the superdiffusive transport to the nature of domain wall perturbation remains an open theoretical problem. 

\subsection{Dynamics of supersymmetric model}
\begin{figure}[h!]
\includegraphics[width=7in]{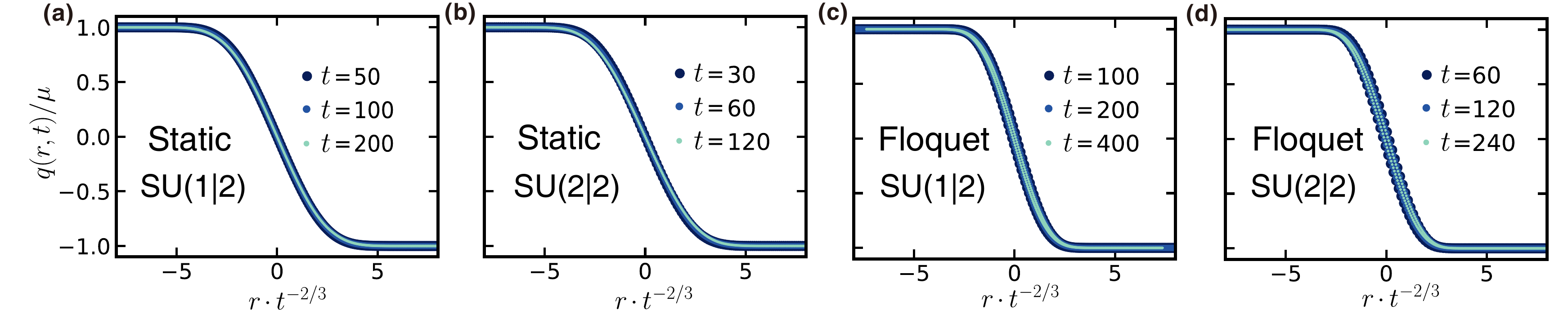}
\centering
\caption{In integrable models with supersymmetry, collapse of the polarization profiles at different times by rescaling the spatial axis with dynamical exponent $z=3/2$. 
}
\label{fig:collapse_supersymmetry}
\end{figure}

In the main text, Fig.~3(f) diagnoses the dynamical exponent $z=3/2$ in the supersymmetric models via total polarization transfer $\mathcal{P}(t)$. 
Here, we show that the collapse of the polarization profile also corroborates this conclusion. 
Same as the symmetric cases (Fig.~1(c) of the main text), we plot the polarization profile as a function of rescaled position $r\cdot t^{-2/3}$, and observe perfect collapse across the entire late-time evolution (Fig.~\ref{fig:collapse_supersymmetry}). 
We note that similar to the Floquet symmetric model, the Floquet supersymmetric models also exhibit a small splitting between even and odd sites, whose size decays with time \cite{ljubotina:2019}. 

\begin{figure}[h!]
\includegraphics[width=2.5in]{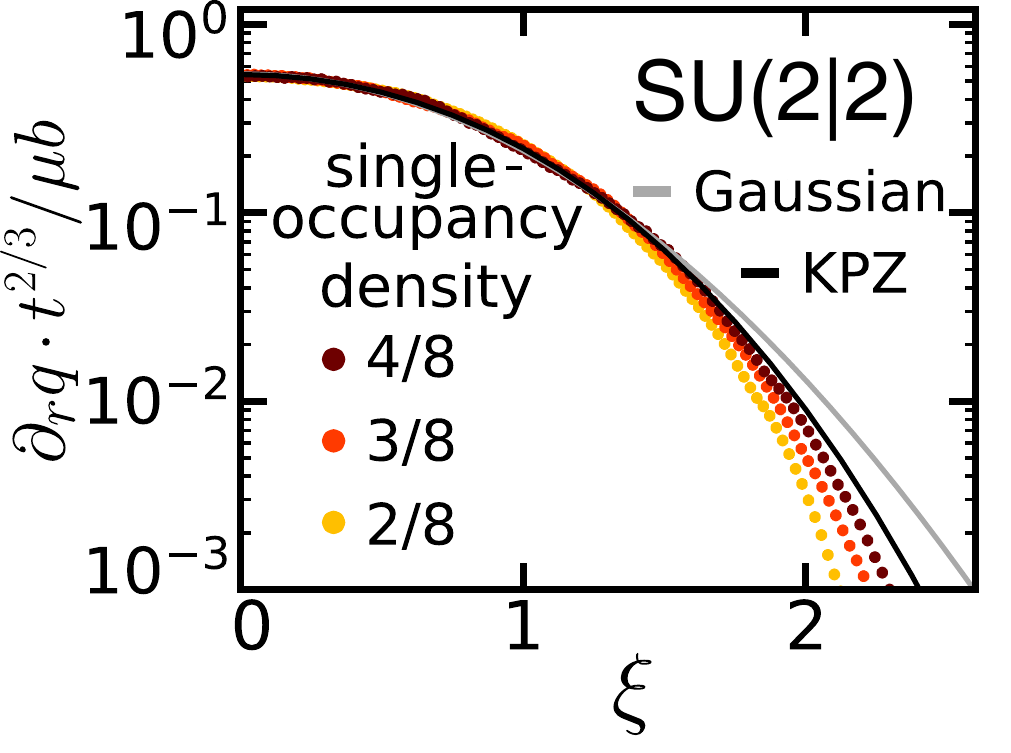}
\centering
\caption{Polarization gradients in an integrable SU(2$|$2) model with varying density of single-occupancy sites. At the same evolution time, the system with smaller single-occupancy density is closer to the KPZ expectation, indicating a faster convergence in time. 
}
\label{fig:SU22_scaling_func}
\end{figure}

For the SU(1$|$2) supersymmetric model, we compare the gradient of the polarization profile against the KPZ scaling function in the main text. 
In particular, while the evidence of KPZ dynamics is less clear at the time our simulation accesses, we observe a clear flow of the polarization gradient towards the KPZ scaling function. 
Moreover, by decreasing the hole density in the initial state, the dynamics exhibit an improved convergence to KPZ universality. 
Here, we perform the similar analysis for the SU(2$|$2) supersymmetric model. 
In particular, the SU(2$|$2) supersymmetric model can be thought of as a special case of the spin-1/2 Fermi-Hubbard model. 
We focus on the transport of the bosonic charge corresponding to the density difference between doublon and vacancy. 
We remark that another Bosonic charge is the spin density, i.e. the density difference between spin-up and spin-down fermions, which exhibit the same superdiffusive transport. 
Analogous to the vacancy density in the SU(1$|$2), we tune the density of single-occupancy site in the SU(2$|$2), and study the convergence of polarization gradient to the KPZ scaling function. 
Crucially, we observe qualitatively the same behavior as in the SU(1$|$2) case: as the single-occupancy density decreases, the dynamics approach to the KPZ expectation at earlier times (Fig.~\ref{fig:SU22_scaling_func}). 

\section{Additional analysis of symmetry breaking and integrability breaking models}
In the main text, we analyzed the dynamics when adding various perturbations to superdiffusive systems. 
Here, we provide more details on the numerics and our extraction of the transport coefficients. 

\subsection{Adding non-zero magnetization}
\begin{figure}[h!]
\includegraphics[width=5.8in]{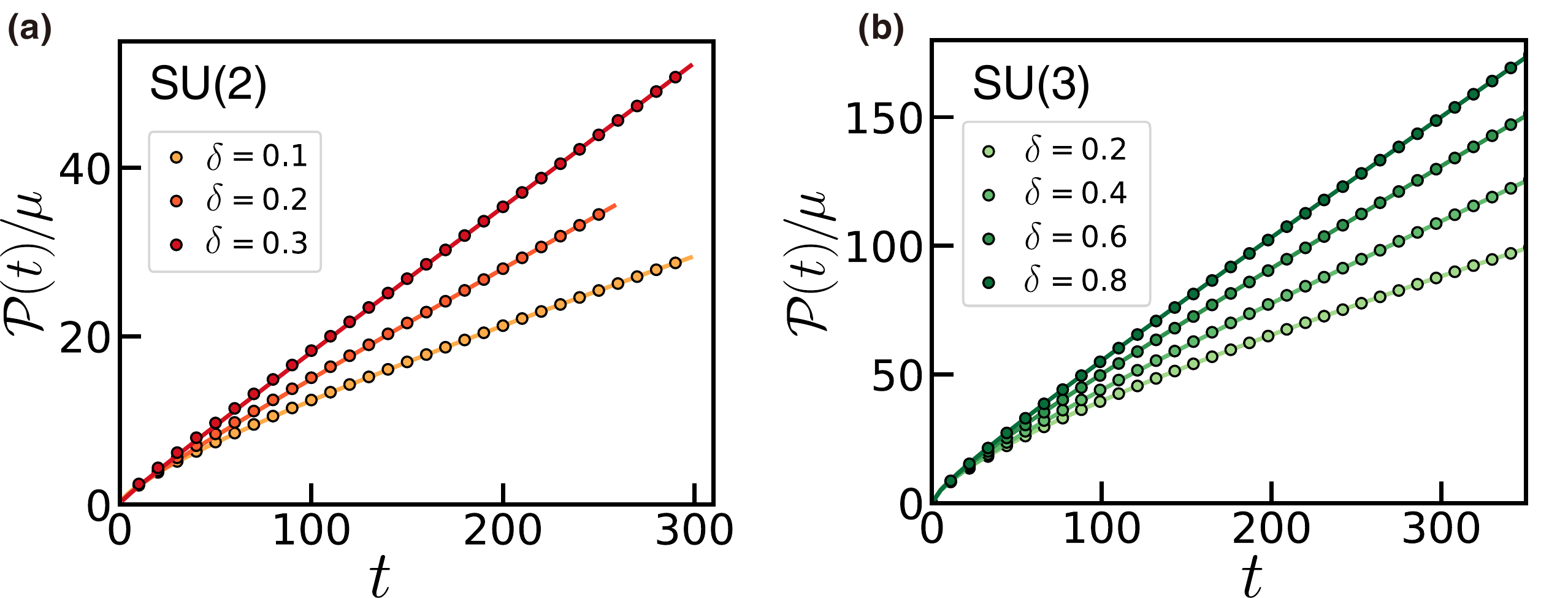}
\centering
\caption{Total population transfer with magnetized initial states. For both (a) the SU(2) model and (b) the SU(3) model, the dynamics exhibits a crossover from the early-time superdiffusive-like transport to the late-time ballistic transport. The data (circles) are well fitted with the functional form $(v^2t^2+At^{\frac{4}{3}})^{\frac{1}{2}}$ (solid lines). 
}
\label{fig:ballistic}
\end{figure}

With a non-zero magnetization in the initial state, the charge will transport ballistically at late enough times. 
However, at short times, the transport should still exhibit some superdiffusive feature and smoothly crossover to the ballistic regime, since intuitively the perturbation will not sharply alter the dynamics. 
This is indeed borne out by our data [Fig.~\ref{fig:ballistic}, for both the SU(2) and the SU(3) models]. 
Moreover, we observe that the larger the magnetic field strength $\delta$, the earlier the crossover to the ballistic regime. 

To extract the velocity $v$ of the ballistic motion, we considering the fitting functional form $(v^2t^2+At^{\frac{4}{3}})^{\frac{1}{2}}$, which takes into account of the crossover from the early-time superdiffusion. 
As shown in Fig.~\ref{fig:ballistic}, such form perfectly fits all data, across different values of $\delta$. 
This can be understood as follows. 
The mean squared displacement $\langle r^2 \rangle\propto t^2\langle h^2 \rangle= t^2\bar{h}^2+t^2\langle (h-\bar{h})^2 \rangle$; the first term is the net magnetization squared $\propto t^2\delta^2$, and the second term is due to the fluctuation of the local magnetization $\propto t^{4/3}$ \cite{gopalakrishnan:2019}. 
In Fig.~2c of the main text, we show the linear dependence of $v$ on $\delta$, corroborating the analytically proposed underlying mechanism of the superdiffusion. 

\subsection{Breaking the symmetry}
\begin{figure}[h!]
\includegraphics[width=6.5in]{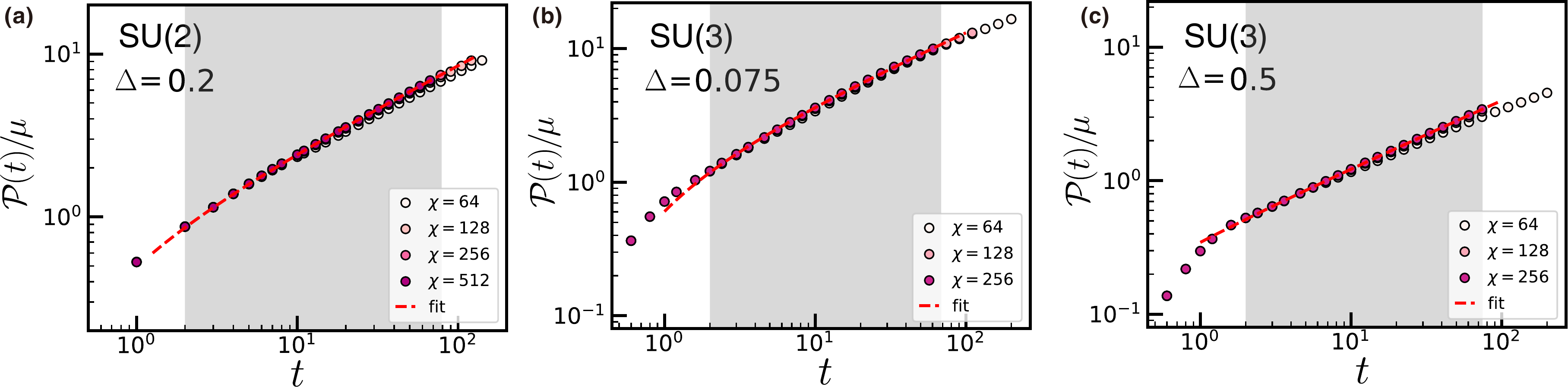}
\centering
\caption{Total polarization transfer under the symmetry-broken Hamiltonian. For both (a) the SU(2) model and (b)(c) the SU(3) model, the dynamics exhibits diffusive transport at late times. As the bond dimension $\chi$ increases, the dynamics become more consistent with diffusion, and the extracted diffusion coefficient slightly flows to a larger value. Dashed line: fit for the data with the largest $\chi$. Shaded region: the time window in which the data is fitted. Since the SU(2) and the SU(3) models are completely unrelated and follow completely different parametrization, one should not directly compare the symmetry breaking strength $\Delta$ between these two different models. 
}
\label{fig:diffusive}
\end{figure}

When breaking the non-Abelian symmetry in the Hamiltonian, one observes diffusive transport at late times. 
However, similar to the case in the above subsection, the early-time dynamics should also exhibit superdiffusive features [Fig.~\ref{fig:diffusive}, for both the SU(2) and the SU(3) models]. 
To extract the diffusion coefficient, we fit the polarization transfer $\mathcal{P}(t)$ with $\sqrt{4Dt/\pi}+C$ at late-times ($t>10/J$). 

Three remarks about the fit are in order. 
First, we allow a constant intercept $C$ in the fitting functional form, which is irrelevant when $t\rightarrow \infty$, but it allows a better fitting (especially for the weak symmetry breaking cases) given finite-time numerical data.
Second, the prefactor $\sqrt{4/\pi}$ is determined by solving the diffusion equation given a domain wall initial state. 
Third, as the bond dimension $\chi$ increases, we observe a robust approach to diffusion at later times, but the extracted diffusion coefficient $D$ also increases (Fig.~2d in the main text). 
Interestingly, for larger symmetry breaking, the extracted $D$ shows a better convergence of $\chi$. 

\subsection{Breaking the integrability}
One may also observe diffusive behavior by breaking the integrability of the Hamiltonian without breaking the SU(2) symmetry by adding next-nearest-neighbour terms of the form described in equation~\eqref{eq:intbreaking}. The numerical simulation shown in the main text (Fig.~2d) is taken for $J_{nnn} = J$. It is necessary to use such a strong perturbation to observe clear diffusion at finite times because the superdiffusive transport is particularly stable (at low order in perturbation theory) to symmetry-preserving perturbations~\cite{denardis:2021}. The next-nearest-neighbour terms in the Hamiltonian are implemented in DMT using SWAP gates, in exactly analogous fashion to their use in TEBD to implement long-range interactions~\cite{schollwock:2011}.

\section{Details of the experimental proposal}

\subsection{Derivation of the SU($N$) Hamiltonian from the bare experimental setup}
In the experimental setup discussed in the main text, the optically trapped alkaline-earth fermions interact via on-site Hubbard coupling. 
Here we show how to obtain the integrable SU($N$) Hamiltonian from the bare Fermi-Hubbard model. 
As a starting point, the spin-$N$ Fermi-Hubbard model is written as:
\begin{equation}
    H^{\mathrm{F-H}} = -t\sum_{i,\alpha}( c^\dag_{i,\alpha}c^{}_{i+1,\alpha}+ c^\dag_{i+1,\alpha}c^{}_{i,\alpha})+\frac{U}{2}\sum_{i} n_i(n_i-1),
\end{equation}
where $t$ is the hopping strength of a fermion between two adjacent sites, $U$ is the on-site repulsive interaction, $n_i$ is the total fermion occupation operator on site $i$, $c^\dag_{i,\alpha}$ and $c_{i,\alpha}$ are the creation and the annihilation operator of species $\alpha$ fermion on site $i$, respectively. 

Assuming $U\gg t$ and the filling factor is $1/N$ (i.e., one fermion per site) the system is in the Mott-insulating phase. 
To the lowest order, the dynamics arise from virtual hopping processes and are generated by the following effective Hamiltonian:
\begin{equation}
\begin{split}
    H^{\mathrm{F-H}} &\approx \frac{2t^2}{U}\sum_{i,\alpha\neq\beta} c^\dag_{i,\alpha}c^{}_{i,\beta}c^\dag_{i+1,\beta}c^{}_{i+1,\alpha}-c^\dag_{i,\alpha}c^{}_{i,\alpha} c^\dag_{i+1,\beta}c^{}_{i+1,\beta}\\
    &= \frac{2t^2}{U}\sum_{i,\alpha\neq\beta} S^{\alpha,\beta}_i \otimes S^{\beta,\alpha}_{i+1}-S^{\alpha,\alpha}_i \otimes S^{\beta,\beta}_{i+1}\\
    &= \frac{2t^2}{U}\sum_{i,\alpha\neq\beta} S^{\alpha,\beta}_i \otimes S^{\beta,\alpha}_{i+1}+\frac{2t^2}{U}\sum_{i,\alpha} S^{\alpha,\alpha}_i \otimes S^{\alpha,\alpha}_{i+1}-\frac{2t^2}{U}\sum_{i} \left( \sum_{\alpha}S^{\alpha,\alpha}_i\right) \otimes \left( \sum_{\beta}S^{\beta,\beta}_{i+1}\right)\\
    &= \frac{2t^2}{U}\sum_{i,\alpha,\beta} S^{\alpha,\beta}_i \otimes S^{\beta,\alpha}_{i+1}-\frac{2t^2}{U}\sum_{i} \mathds{1}_i\otimes\mathds{1}_{i+1},
\end{split}
\end{equation}
which is exactly the SU($N$) integrable Hamiltonian (Eq.~4 in the main text) up to a constant energy shift. 

A few remarks are in order. 
First, while the effective Hamiltonian should certainly inherit the SU($N$) symmetry from the bare Hamiltonian, the same is not necessarily true for integrability. 
Indeed, when the filling factor is not $1/N$ [or $(N-1)/N$], the effective Hamiltonian is not integrable. 
Second, the next-order correction to the effective Hamiltonian is of order $\frac{t^4}{U^3}$, which in principle will slightly break the integrability and alter the very-long-time transport dynamics. 

\subsection{Initial state preparation}
Besides the simulation of the correct Hamiltonian, it is crucial to also accurately prepare a near-infinite temperature domain wall initial state.
In current optical lattice experiments, the main challenge in preparing such states arises from the need to controllably add entropy to the system \emph{after cooling the motional degrees of freedom}.
In past experiments~\cite{wei:2021}, this was accomplished by applying a combination of global rotations to control the spin population, and single-site rotations to dephase individual spins with respect to one another; averaging over different single-site rotations yields the mixed initial state of interest.
However, when moving from spin-1/2 to larger spin systems (necessary to host SU($N$) symmetric models), this protocol is not suitable to prepare arbitrary spin population distributions, since the population distribution is only changed when applying the global rotation of the spins.

\begin{figure}
    \centering
    \includegraphics[width=0.8\textwidth]{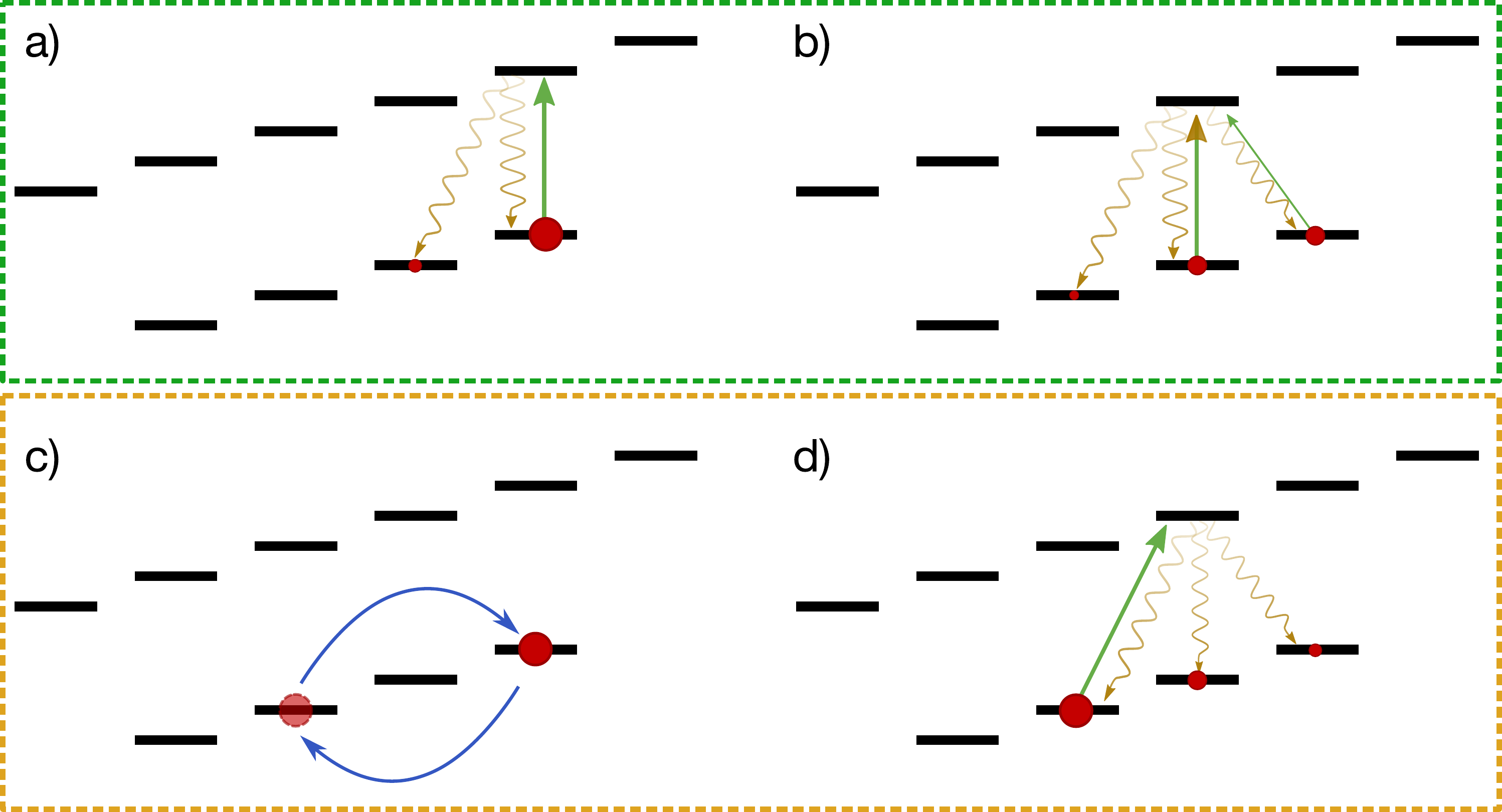}
    \caption{Two different approaches towards preparing arbitrary population distributions. 
    {\bf a)} Starting from the $|F,F\rangle$ state in the ground state manifold, $\hat{z}$ polarized laser can be used to move population to $|F,F-1\rangle$. 
    {\bf b)} To continue this process toward lower $m_F$ states, one can again apply a $\hat{z}$ polarized laer. However, this would lead to additional leakage back to a higher $m_F$ state. To counteract this effect, an additional $\sigma^-$ polarized laser field should be applied to prevent leakage.
    {\bf c)} Another approach to prepare a spin population distribution begins by moving the entire population from the $|F,F\rangle$ into the lowest $|F,f_{\min}\rangle$ of interest.
    {\bf d)} From $|F,f_{\min}\rangle$, the population distribution can be built by iteratively applying a $\sigma^+$ polarized laser to move polarization from $|F,f\rangle$ to $|F,f+1\rangle$ and $|F,f+2\rangle$.
    }
    \label{fig:optical}
\end{figure}

To circumvent this, we propose a different scheme, which utilizes the flexibility provided by manipulating the atoms \emph{before} loading them into the optical lattice. 
Crucially, this allows us to leverage optical pumping techniques using the excited hyperfine manifold to move population density across the different spin levels of interest, Fig.~\ref{fig:optical}. 
Let us consider two possible different implementations of this idea:
\begin{itemize}
    \item Starting with unit population in the largest $m_F$ state of the manifold $|F,m_F=F\rangle$, a $\hat{z}$ polarized laser field is shined, exciting the system to the corresponding hyperfine state in the excited manifold, $|F+1,F\rangle$.
    Spontaneous emission then leads to decay back to the ground state manifold, into $|F,F\rangle$ and $|F,F-1\rangle$, with a branching ratio dictated by Clebsch-Gordan coefficients. By timing the laser duration appropriately, this enables the arbitrary population transfer from the $|F,F\rangle$ to the $|F,F-1\rangle$ state, Fig.~\ref{fig:optical}a.
    In order to transfer from $|F,f \rangle$ to $|F,f-1 \rangle$, we can follow the same strategy, using a laser field to excite to $|F+1,f \rangle$, and then use the spontaneous decay to change the population distribution.
    However, since there can also be decay into $|F, f+1 \rangle$, this method requires an additional laser field to overcome this leakage. The simplest approach is to use a $\sigma^-$ polarized laser to excite from $|F, f+1 \rangle$ back into  $|F+1, f \rangle$ and counteract this leakage effect, Fig.~\ref{fig:optical}b.
    Once again, by controlling the relative power between the two laser fields, as well as their timing, one can transfer a precise amount of population from $|F, f \rangle$ to $|F, f-1 \rangle$. Doing this procedure repeatedly enables the preparation of an arbitrary spin population distribution.\\
    \item While conceptually simple, the above protocol may prove difficult owing to the need of two different laser fields with different polarizations.
    We now discuss a different protocol which, after an initial coherent population transfer, only requires a single laser at any point of time.
    Starting again with unit population in the largest $m_F$ state of the manifold $|F,F\rangle$, we begin by performing a coherent rotation between $|F,F\rangle$ and the lowest hyperfine state we wish to have finite population, say $|F, f_{\min}\rangle$, Fig.~\ref{fig:optical}c.
    With unit population in $|F, f_{\min}\rangle$, we can now apply a $\sigma^+$ polarized laser to excite the system into $|F+1, f_{\min}+1\rangle$. Subsequent spontaneous decay will transfer the polarization into $\{ |F, f_{\min}\rangle, |F,f_{\min}+1\rangle, |F,f_{\min}+2 \rangle\}$. By doing this process for the correct amount of time, the population in $|F, f_{\min}\rangle$ can be set to the desired value.
    Critically, this procedure can be performed iteratively---at each step the population in level $|F,f\rangle$ can be set by exciting any extra population into $|F+1,f+1\rangle$ and moving it into higher hyperfine states, $|F,f+1\rangle$ and $|F,f+2\rangle$, Fig.~\ref{fig:optical}d.
    As long as the upper manifold has equal or larger total angular momentum, this process can be carried out until reaching back to the $|F,F\rangle$ state, using an excitation from $|F,F-1\rangle$ to $|F+1,F\rangle$ to set its population.
\end{itemize}

\begin{figure}
    \centering
    \includegraphics[width=0.8\textwidth]{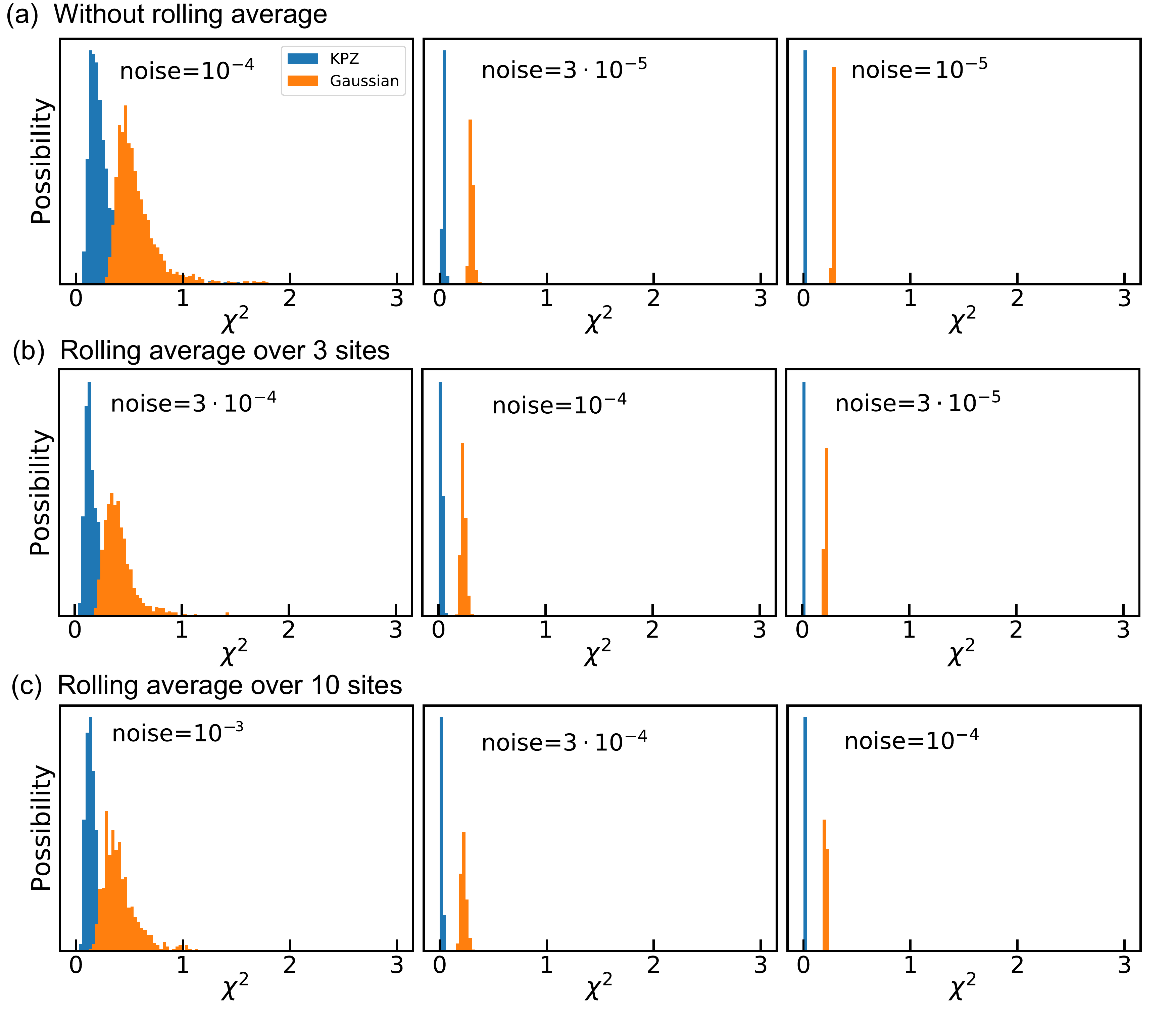}
    \caption{$\chi^2$ analysis of the confidence to confirm the KPZ dynamics. By adding random noise on the data from DMT numerics, we can mimic experimental data. For each noise realization drawn from a normal distribution, we fit the noisy data with both the KPZ and the Gaussian expectations, and obtain the corresponding $\chi^2$. 
    The histogram of $\chi^2$ indicates the confidence of whether the data agrees better with KPZ or Gaussian expectation. 
    As the size of noise decreases, $\chi^2$ corresponding to KPZ approaches to zero and is separated from $\chi^2$ corresponding to Gaussian fit. 
    Performing rolling average of polarization profile over $n$ consecutive sites can effectively reduce the effect of noise roughly by a factor of $n$. The numerics are performed on the integrable SU(3) model with $\mu=0.1$, and we choose to fit the polarization profile at $t=52/J$. 
    }
    \label{fig:chi_sq_all}
\end{figure}

\begin{figure}
    \centering
    \includegraphics[width=0.45\textwidth]{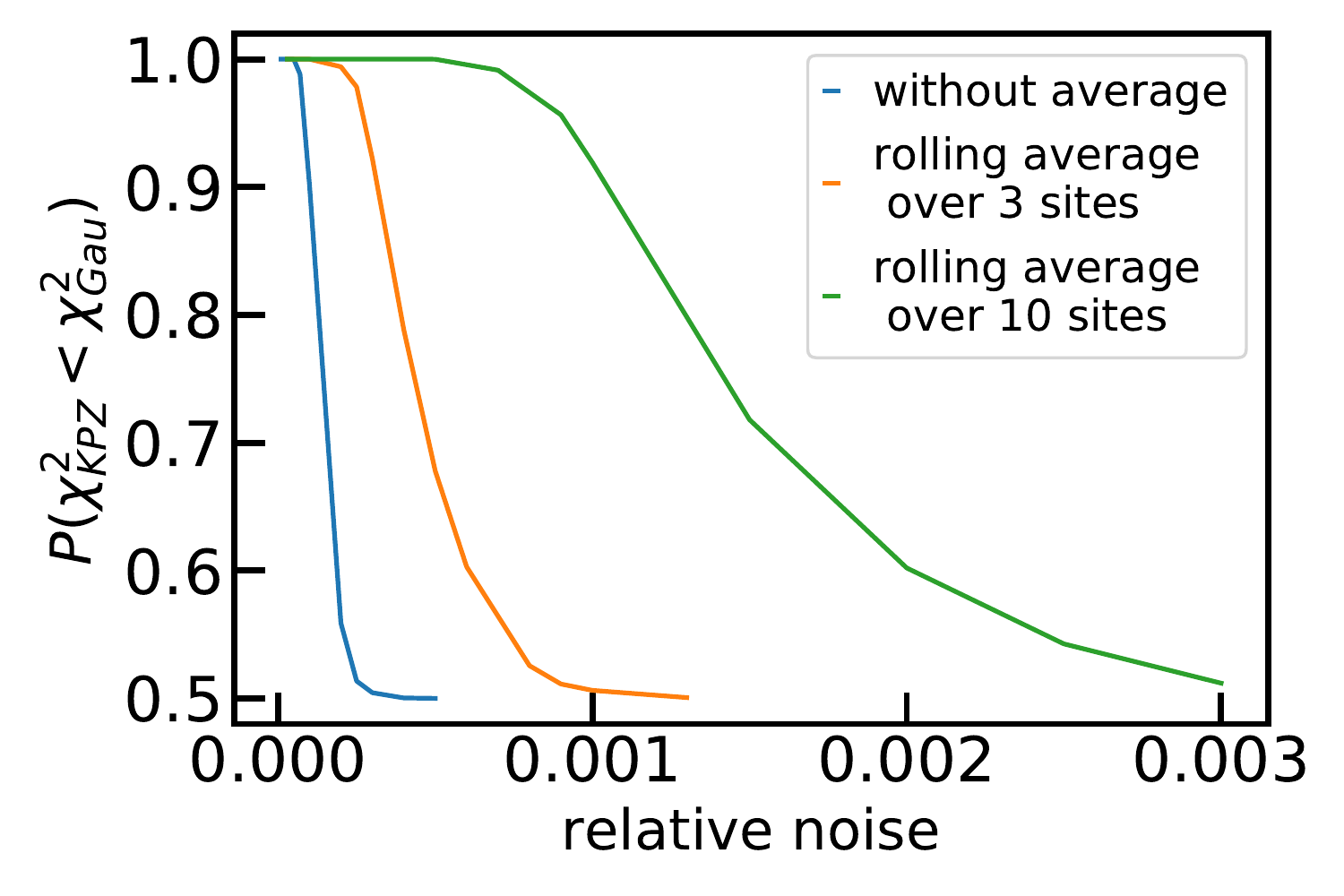}
    \caption{$\chi^2$ analysis to confirm KPZ dynamics and to exclude Gaussian expectation. To quantify the confidence with which we can conclude KPZ fits the data better, we compute the possibility of obtaining a smaller $\chi^2$ in the KPZ fit than in the Gaussian fit. Without performing a rolling average, when the relative noise amplitude is below $10^{-4}$, it becomes almost certain that the KPZ expectation fits the data better, suggesting that the experiment can conclusively confirm the KPZ expectation at such noise level. 
    By further performing a rolling average over $n$ sites, we can increase the minimum relative noise required by roughly $n$ times. 
    }
    \label{fig:chi_sq_possibility}
\end{figure}

\begin{figure}
    \centering
    \includegraphics[width=0.45\textwidth]{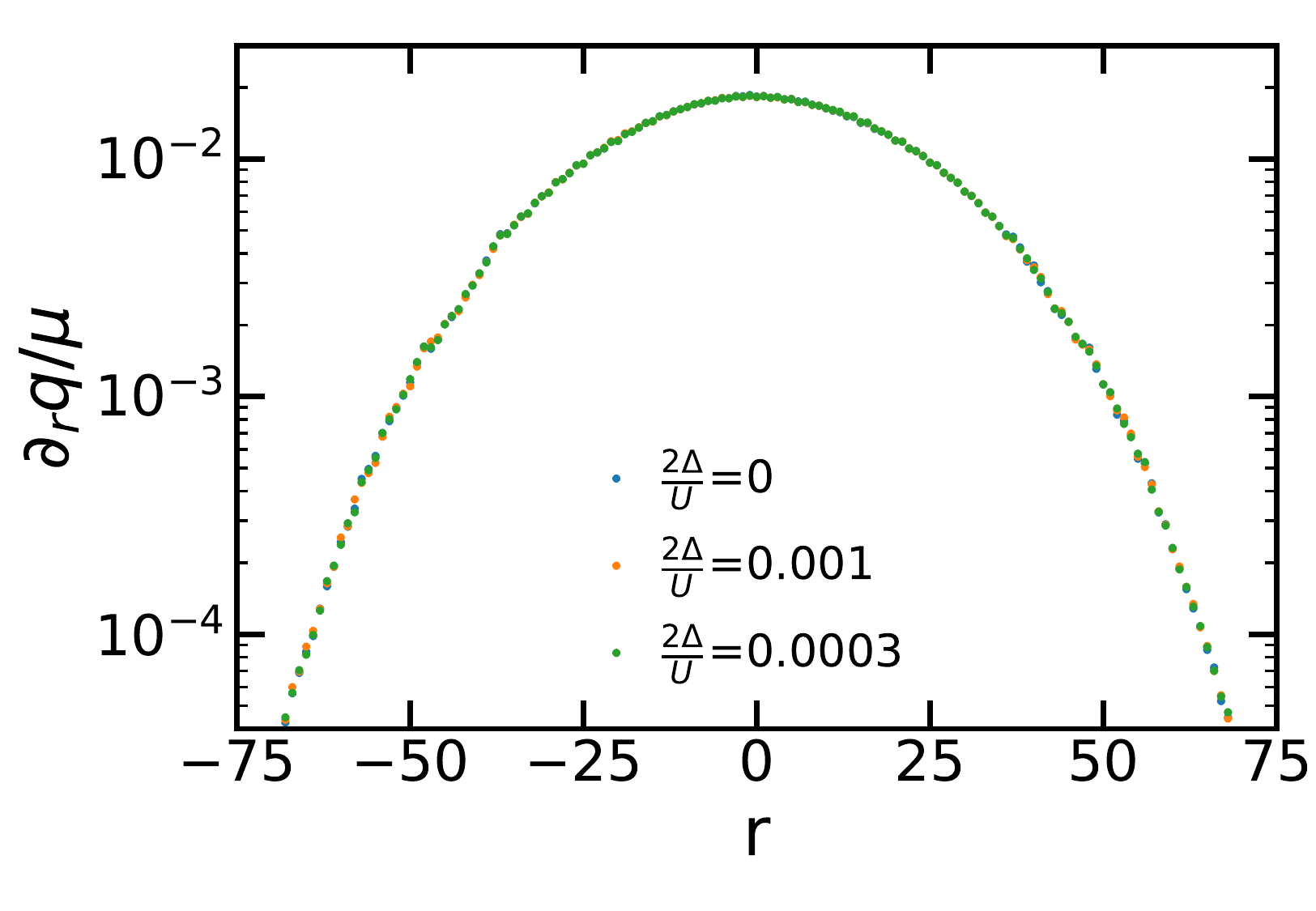}
    \caption{Influence of a quadratic inhomogeneity of the local onsite energy. Evolving under the estimated size of the inhomogeneity in experiment, the spatial gradient of the polarization profile remains almost the same, and does not affect the observation of the KPZ dynamics. The numerics are performed for the integrable SU(3) model. 
    }
    \label{fig:quad}
\end{figure}

\subsection{Distinguishing KPZ from Gaussian scaling function}

Since the distinction between the KPZ and  Gaussian scaling functions is only evident in the tails of the distributions, it requires measuring the system's scaling function to quite a small uncertainty.
In this section, we quantify the requisite noise floor of an experiment whose goal is to distinguish (e.g.~using a $\chi^2$ test) between these two possibilities.
Crucially, this informs us, given a choice of time ($t\sim 50/J$) and system size ($L=150$), the required experimental uncertainty.

Starting with the numerical calculation of the scaling function from DMT (Fig.~3a in the main text), we mimic the effect of experimental uncertainty by adding random Gaussian noise of constant strength to the measured spin.
For each instance of random noise, we can compute the $\chi^2$ of the fits to the Gaussian function and the KPZ scaling function. 
By repeating this analysis for different noise instances, we obtain the distribution of $\chi^2$. 
Crucially, when the relative noise $\lesssim 10^{-4}$, the two distributions are distinguishble and we can confidently exclude the Gaussian scaling function as the correct description of the scaling data, Fig.~\ref{fig:chi_sq_all}.
In Fig.~\ref{fig:chi_sq_possibility}, we compute the probability of the KPZ $\chi^2$ value being smaller than the Gaussian $\chi^2$---at a relative noise of $10^{-4}$ is it approximately $85\%$.

At the same time, we find that performing a rolling average over $\bar{n}$ points (thus effectively reducing the error strength by $\bar{n}$), enables to differentiate the two distributions at a larger of relative error---using $\bar{n}=10$, a relative error of $\sim 10^{-3}$ already distinguishes the two hypothesis, Figs.~\ref{fig:chi_sq_all} and \ref{fig:chi_sq_possibility}.

\subsection{Estimating experimental runtime}
From the analysis above, we require relative magnetization uncertainty on the order of $\delta \bar{m}/\mu \sim 10^{-4}$ to determine whether the observed magnetization dynamics falls under the KPZ or Gaussian universality class.
Leveraging the rolling average over $\bar{n}$ sites reduces this requirement to $\delta \bar{m}/\mu \sim 10^{-3}$.
In order to operate within the small domain-wall regime where the gradient of the polarization accurately captures the two-point correlation function, we need the domain wall size to be at most $\mu \lesssim 0.2$, leading to an absolute uncertainty in the magnetization error of $\delta \bar{m} \sim 2\times 10^{-4}$.
Crucially, a projective measurement of a single spin with $N$ levels follows a multinomial distribution with variance $1 / N$ thus requiring a total of $N_{\mathrm{samples}} > \frac{1}{N\delta \bar{m}^2}  = 2.5\times 10^6 $ spin measurements.
As we demonstrate below, this is within the realm of current experimental capabilities.

For instance, a two-dimensional cavity-enhanced $813~\mathrm{nm}$ optical lattice can be loaded with a Mott-insulator of $^{87}$Sr atoms as large as 300 sites in diameter~\cite{park:2021} then subsequently divided into $m_x = 250$ parallel chains, $L = 150$ sites in length.
Assuming an on-site interaction energy of $U \approx 3.2~\mathrm{kHz}$ and a tunneling rate along the 1D-chains of $t_y\approx300~\mathrm{Hz}$ gives a coupling strength of $J = 2t_y^2/U \approx 56~\mathrm{Hz}$.
Decoherence rates in current implementations of cavity-buildup optical lattices have been shown to limited by heating out of the ground band due to intra-cavity intensity noise at the $T_\mathrm{Heating} \approx 80~\mathrm{s}$ level for relevant lattice depths~\cite{park:2021}.
However, with moderate improvements in intrinsic laser noise as in Ref.~\cite{eckner:2021}, coherence is limited by single atom loss due to finite vacuum pressure at a time of $T_{\text{atom loss}} > 180~\mathrm{s}$~\cite{park:2021}.
Post-selection on chains with fewer than $L$ atoms can remove such effects from the analysis while increasing the required number of iterations by a factor of $\epsilon = \exp(L t / T_{\text{atom loss}}) \sim 2.1$, where $t\sim 50/J$ is the duration of the coherent evolution. 
Thus a single experimental cycle provides $ N_{\text{samples per exp}} = 2\times m_x / \epsilon \approx 2.4\times 10^2$ spin measurements (where $2$ comes from averaging both the two sides of the domain wall). 
As a result, one requires $N_{\mathrm{exp}} = N_{\mathrm{samples}} / N_{\text{samples per exp}} = 1.1\times 10^{4}$.
Because each experimental cycle requires $\tau \sim 10~\mathrm{s}$ to run~\cite{sonderhouse:2020}, we estimate a total averaging time of $N_{\mathrm{exp}} \tau \sim 1.1\times 10^5~\mathrm{s} \sim 30~\mathrm{hours}$ is required.

Let us finally remark that in practice, the lattice may not be entirely homogeneous. 
The shape of the Gaussian beam leads to a small quadradic inhomogeneity on the local onsite energy $V(r) = \Delta r^2 n_r$, which modifies the hopping between nearby sites to be $J(r) \sim J\left[1 + (\frac{2\Delta}{U} r)^2\right]^{-1}$.
Nevertheless, using the experimentally demonstrated $\Delta = 1~\mathrm{Hz}$, we observe no effect on the system's dynamics within the timescales proposed, Fig.~\ref{fig:quad}.

\bibliography{KPZ}